\documentclass[prodmode,acmtochi]{acmsmalladapted} 
  
\ccsdesc[500]{Human-centered computing~HCI theory, concepts and models}

\makeatletter
\def\url@leostyle{%
  \@ifundefined{selectfont}{\def\UrlFont{\sf}}{\def\UrlFont{\small\bf\ttfamily}}}
\makeatother
 \setcopyright{none}
\usepackage{graphicx}
\usepackage[procnames]{listings}
\usepackage{siunitx}
\usepackage{placeins}
\usepackage{balance}  % to better equalize the last page
\usepackage{graphics} % for EPS, load graphicx instead 
\usepackage{todonotes}
\usepackage{float}
\let\xtodo\todo
\renewcommand{\todo}[1]{\xtodo[inline,color=black!5]{#1}}

\usepackage[printonlyused, nolist]{acronym}

\usepackage{tabularx}
\usepackage{multirow} 
\usepackage{array}

\usepackage{subfigure}

\makeatletter
\def\url@leostyle{%
  \@ifundefined{selectfont}{\def\UrlFont{\sf}}{\def\UrlFont{\small\bf\ttfamily}}}
\makeatother
\graphicspath{{simon_gp/}{matlab_plots/}{Figures2/}{Figures/}}

\def\pprw{8.5in}
\def\pprh{11in}

\definecolor{linkColor}{RGB}{6,125,233}

\begin{document}

\definecolor{keywords}{RGB}{255,0,90}
\definecolor{comments}{RGB}{0,0,113}
\definecolor{red}{RGB}{160,0,0}
\definecolor{green}{RGB}{0,150,0}
 
\lstset{language=Python, 
        basicstyle=\ttfamily\small, 
        keywordstyle=\color{keywords},
        commentstyle=\color{comments},
        stringstyle=\color{red},
        showstringspaces=false,
        identifierstyle=\color{green},
        procnamekeys={def,class}}

\title{Forward and Inverse models in HCI:Physical simulation and deep learning for inferring 3D finger pose}

\author{Roderick Murray-Smith
\affil{School of Computing Science, University of Glasgow}
John H. Williamson
\affil{School of Computing Science, University of Glasgow}
Andrew Ramsay
\affil{School of Computing Science, University of Glasgow}
Francesco Tonolini
\affil{School of Computing Science, University of Glasgow}
Simon Rogers
\affil{School of Computing Science, University of Glasgow}
Antoine Loriette
\affil{School of Computing Science, University of Glasgow}
}

\begin{abstract}
We outline the role of forward and inverse modelling approaches in the design of human--computer interaction systems. Causal, forward models tend to be easier to specify and simulate, but HCI requires solutions of the inverse problem. We infer finger 3D position $(x,y,z)$ and pose (pitch and yaw) on a mobile device using capacitive sensors which can sense the finger up to 5cm above the screen. We use machine learning to develop data-driven models to infer position, pose and sensor readings, based on training data from: 1. data generated by robots, 2. data from electrostatic simulators  3. human-generated data. 
Machine learned emulation is used to accelerate the electrostatic simulation performance by a factor of millions. We combine a Conditional Variational Autoencoder with domain expertise/models experimentally collected data. We compare forward and inverse model approaches to direct inference of finger pose. The combination gives the most accurate reported results on inferring 3D position and pose with a capacitive sensor on a mobile device.
\end{abstract}

\maketitle

%\category{H.5.2}{User Interfaces}{Input devices and strategies}. 
%\keywords{Forward model, inverse model, 3D touch, machine learning, electrostatic simulation}

\section{Introduction}
Interactive systems must be able to sense and interpret human actions to infer their intentions. Commercial device developers and HCI researchers continually explore the use of novel sensors to enable novel forms of interaction. However, we currently lack a coherent, consistent framework for characterising this process with incrementally improving precision for different sensors and different human behaviours. Common practice today tends to be to hand-craft features and associated thresholds for specific use-cases. This can be very time-consuming, and unpredictable in outcome, especially as the dimension or variability of the sensors increase. Furthermore, the thresholds for one application might not be appropriate for another (e.g. touch typing vs continuous gestures).

We argue that for the field to make consistent incremental progress, we need a more general, formal framework for characterisation of the pathway from human intent to sensor state. This pathway can include formal, computational models of human elements such as cognition and physiological processes, as well as purely technical elements such as the characterisation of the physical processes of the sensor.\footnote{Eventually this forward model could include the anticipated impact of the feedback from the interface on the forward process.}

\subsection{Forward models and inverse problems in HCI}
As a foundation for more predictable incremental progress in HCI, let us take inspiration from traditional scientific methods. Scientific theories allow us to make predictions. If we have a complete mathematical model of a physical system we can predict the outcome of measurements of states of that system. The {\it forward problem} is this problem of predicting the results of measurements. The {\it inverse problem} is the problem of using measurements to infer values of parameters that characterise the system or its inputs \cite{Tar05}. In many cases, scientists can simulate a system better than they can effectively observe it.

\begin{figure}[htb]
    \centering
    \includegraphics[width=0.9\linewidth]{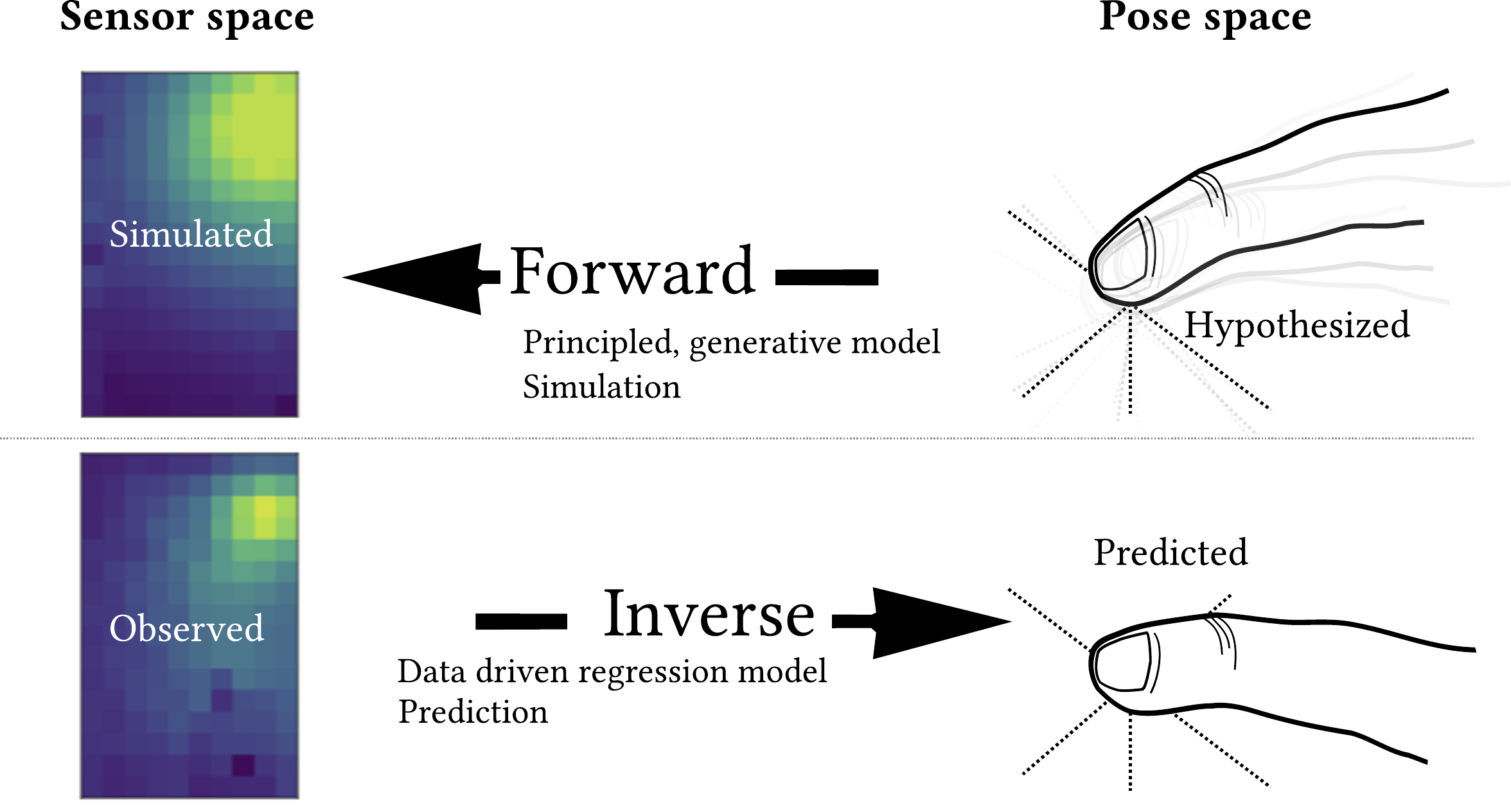}
    \caption{Forward and inverse models in a finger sensing task}
    \label{fig:fwd_inv_p1}
\end{figure}
\begin{figure}[htb]
    \centerline{\includegraphics[width=0.8\linewidth]{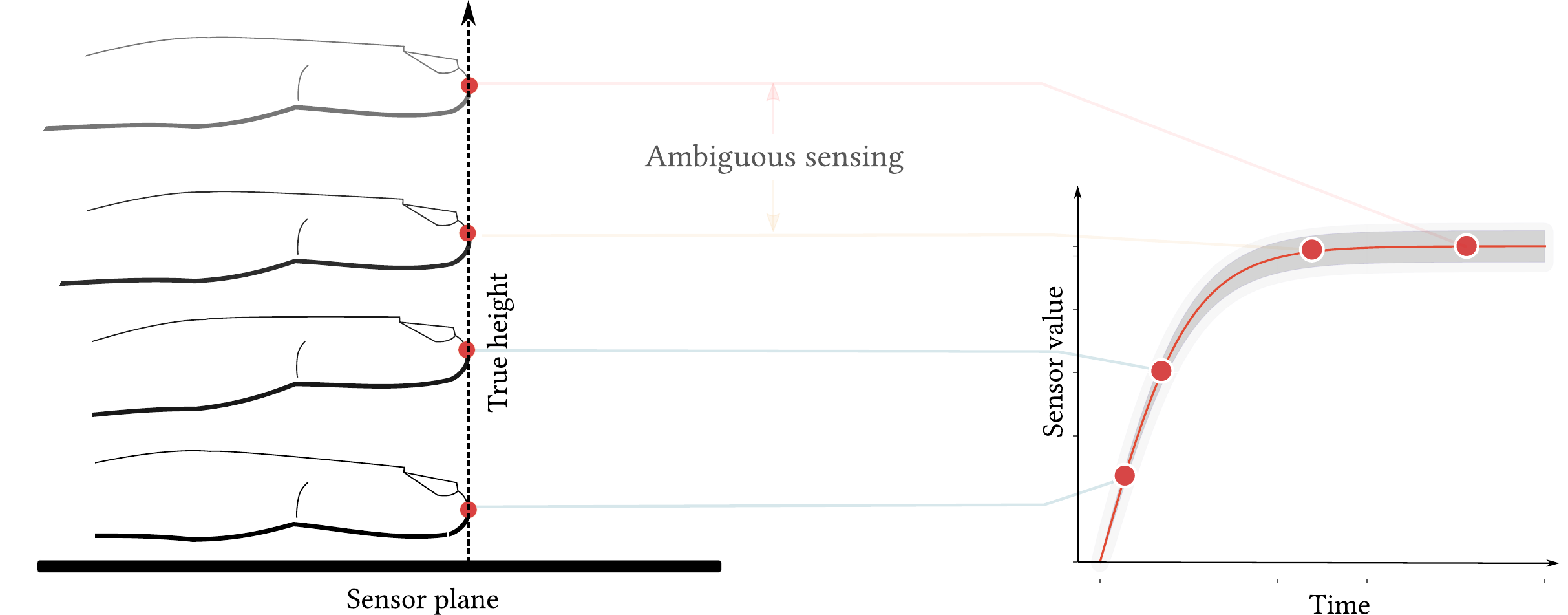}}
    \caption{A saturation effect is one potential cause of ill-posedness. Distinguished world states (finger poses) are projected to identical sensor states.}
    \label{fig:saturation}
\end{figure}

The forward problem for many systems of practical interest tends to have a unique solution, but the inverse problem does not.\footnote{If our model parameters, structures or inputs are uncertain, then we have a distribution of solutions, but for each sample of the uncertain variables we can generate a deterministic prediction.} E.g. take something as simple as a nonlinear saturation effect in the forward mode, as shown in Figure~\ref{fig:saturation}. We can predict the forward values with precision, but there are infinitely many possible solutions to the inverse problem in the saturated areas. This means that solution of inverse problems requires explicit use of {\it a priori} information about the system, and careful consideration of uncertainty in the data. 

We propose a {\it dual approach} where we model both the forward and the inverse problems, and fuse them consistently via a probabilistic framework. The core idea is that there are two types of uncertainty that pervades interaction: {\it epistemic} uncertainty where a lack of knowledge means we unsure as to whether a model of the user, sensor or world is valid, and {\bf aleatoric} uncertainty where measurements are noisy and subject to random variation \cite{Oha04}.  

First-principles models are an attractive way of approaching HCI problems but it is hard for simple, mathematically elegant models to represent messy human behaviour, and the complexities of practical sensing processes for mass-production devices. Data driven approaches have the advantage that they are precisely tailored to specific real interactions, but struggle to generalise robustly. Therefore in this paper we consider a construction which divides up the problem of modelling interaction into two streams:

{\bf Forward}, first-principles models of interaction derived from physics, physiology and psychology, where parameters are not fully known, with {\it epistemic} uncertainty and unique solutions, e.g. implementing an executable simulation of what sensor vector we would observe for a given finger pose.

{\bf Inverse}, data-driven models of interaction, learned from observed interaction with machine learning, with {\it aleatoric} uncertainty and non-unique solutions, e.g. training a regression model to predict pose from a touch sensor vector.

This splits the modelling task into two parts with complementary strengths, bringing generalisable, testable simulation models into the forward construction and powerful machine learned predictive models into the inverse construction, supported by the combination of data-driven and theory-driven components of the forward model. 

We hope to convince the reader that this is both an intellectually appealing as well as a practically useful way of partitioning the analysis of the interaction problem, and using experimental data we will provide a concrete demonstration that probabilistic fusion of forward and inverse models leads to performance which exceeds the performance of directly learning an inverse mapping.

\subsection{Probabilistic approaches}
The approach to modelling in this paper is probabilistic. We are performing inference on the possible human poses, conditioned on the sensed data, and we use machine learning tools which can sample from the distributions learned. \cite{mackay2003information} highlights that probability calculations are either {\it forward probabilities} or {\it inverse probabilities}, where forward probabilities involve a generative mathematical model that describes a process that gives rise to some data. Inverse probabilities also involve a generative process, but instead of computing the probability distribution of a quantity produced by the process, we compute the conditional probability of one or more of the unobserved variables, given the observed ones. This requires the use of Bayes' rule, which is the formal way of describing this process of {\it statistical inversion}. 
We assume there is a unknown quantity $x$ and we observe some noisy measurements $y_i$. The goal of the statistical inversion is to estimate the unknown quantity $x$, i.e. compute the conditional {\it posterior distribution} of the quantity $x$ given the measurements $y$,
\begin{equation}
p(x| y) = \frac{p(y| x)p(x)}{p(y)}.
\label{eq:Bayes}
\end{equation}
Equation~\ref{eq:Bayes}  includes a probability $p(y|x)$  (the `likelihood' of $y$, given $x$) which represents the `forward model' or `measurement model' which describes the causal but inaccurate or noisy relationship between the true parameter $x$ and the observation $y$. The prior distribution $p(x)$ representing our subjective uncertainty about the value of $x$ before we saw the data $y$. It can also be presented as $p(x|y) \propto p(y|x)p(x)$, where the normalisation constant $p(y) = \int p(y|x)p(x) dx$. In principle, this formulation permits the application of the standard computational tools of Bayesian inference from {\it maximum a posteriori} (MAP) methods to Markov-Chain Monte Carlo (MCMC) numerical integration techniques, e.g. \cite{TurPauMul19,sarkka2013bayesian}. However, for real-time use, such methods are too computationally intensive for many applications, so we explore the use of more computationally efficient approximations, using machine learning tools.

\subsection{Inferring finger pose from capacitive sensor inputs}
The ability to sense finger position and pose accurately a distance from the device screen would allow designers to create novel interaction styles, and researchers to better track, analyse and  understand human touch behaviour. Progress in design of capacitive screen technology has led to the ability to sense the user's fingers up to several centimetres above the screen. However, the inference of position and pose is  a classic example of an ill-posed inverse problem, given only the readings from the two-dimensional capacitive sensor pads, making the solutions inherently uncertain.

We have chosen to illustrate the role of forward and inverse models in HCI with the finger pose inference problem, as shown in Figure~\ref{fig:fwd_inv_p1}, as it is a realistic, topical problem, where the physical theory of the associated system is well understood, but the sensors used in a practical mobile device have significant limitations, where there are non unique mappings in the inverse problem and finally, we can potentially easily augment the sensor model with models of human touch movement such as \cite{Oul18a}.

\subsection{Paper Overview} 

This paper addresses several general themes relevant to Computational HCI:
\begin{enumerate}
    \item \noindent {\bf Inverse models in HCI methodology:} We propose that the forward/inverse modelling approach widely used in other areas of science should be considered more carefully as a formal basis for understanding the design of interactive systems.
    \item \noindent {\bf Machine learning:}
We use probabilistic, Deep CVAE neural networks to learn to predict the position and pose of a finger above the screen on a prototype industrial device, comparing the performance with that provided by the sensor's in-built API. We test the use of machine learned models in both {\it forward, causal models} and {\it inverse, regression} models.

    \item \noindent {\bf Empirical models as approximations of computationally complex first-principles models:} The electrostatic models used to generate our simulated data are too computationally-intensive to use in real-time on mobile technology, but we demonstrate that if we generate large amounts of training data using them, we can create a flexible black-box model which accelerates the simulation such that it can be used in real-time, and by this integration of the models into deep network structures, they can be enhanced by experimental data, to be calibrated to the real observed behaviour.
\end{enumerate}
and applies the methods to the concrete problem of {\bf 3D pose inference for capacitive sensing}, where it uses a novel capacitive sensor, and looks at  {\bf multiple approaches to generating training data}: We tested three approaches to data acquisition: 1. data generated by robots,
2. data from electrostatic simulators and
3. instrumented human-generated data.

\section{Background}
\subsection{Forward/inverse models in HCI}
How does this relate to the study of Human--Computer Interaction? Forward models of human or sensor behaviour tend to be more straightforward to specify theoretically and to acquire empirical data from their associated behaviour, and as described above, they more frequently have a unique mapping. Interface designers, however, if they want the computer to be able to respond to the human behaviour,  need inverse models which can let them determine the intention behind the sensed observations. These inverse problems often have non-unique solutions, and are based on incomplete and noisy observations. E.g., we can specify models of human physiology relating to movements of the arm and hand, and gradually improve their fidelity via theoretical insight, a range of experimental techniques and high-fidelity sensors in carefully controlled lab settings. Advanced sensors like those used in Brain Computer Interfaces explicitly include forward models of the skull to infer the brain state from EEG sensors, but the same principle can be applied in all interactive systems. The challenge for the interface designer is, however, to infer from some possibly cheap and low fidelity sensor information which human intentions were associated with movements of their body, which led to the series of sensor readings. Particle filters also use forward and inverse models, and have been used in a range of computer vision \cite{BlaIsa97} and HCI applications \cite{BlaJep98,RogWilSte11}.

\subsection{Inverse problems in input systems}
Similarly, on the engineering side, for a user-interface sensor input system we can often predict the behaviour of the forward model of the system from knowledge of the technical specifications of the series of components used. However, it is difficult to predict whether these will be sensitive enough to be good enough for the interaction requirements, without either building the system and testing it, or by finding inverse models which let us predict the expected accuracy of the resulting system, for typical inputs. The ability to create such a model computationally, in advance of construction would be a useful tool in the design stage of a project. 
\subsection{Inverse problems in human behaviour}
Collaborative efforts to generate executable computational models of human behaviour include \cite{DelAndArn07,ManRei12,HesBroHus11}. Applications of such models in HCI include \cite{BacPalOul15}\cite{Bac16}\cite{Oul18a}. \cite{KanAthHow18} introduced the use of computationally efficient Approximate Bayesian Computation (ABC) approaches to solve inverse problems in HCI. Their approach did not specify an explicit forward model {\it a priori}, but rather learned a forward model for human cognitive processes which would optimise a plausible cost function.

\subsection{Practical computational models}
 However, in many cases the scientific knowledge and models will be locked into legacy code or commercially confidential software, such that the forward models are essentially black-box models which can be executed to generate data, but cannot be conveniently interrogated internally. The availability of flexible algorithmically differentiable models which can learn from data, such as those used in deep learning environments such as Tensorflow \cite{tensorflow2015-whitepaper}, means that if we can stimulate the forward model sufficiently to be able to machine learn a `clone' of it, we also have an analytically differentiable model.

Flexible statistical models have been used as {\bf black-box models of complex simulations} to create more efficient representations of computationally complex simulators \cite{SacWelTob89,KenOHa01,ConGosOak09}. This requires initial simulation effort to generate training data for a machine learning solution, which can then run more rapidly than the original simulations -- the ML-based approach can be viewed as a glorified lookup table which performs inference between the observations to avoid exponential explosions of required storage. We can represent the simulator in the form of a function $y = f(x)$. Each run of the simulator is defined to be the process of producing one set of outputs $y$ for one particular input configuration $x$. We assume that the simulator is deterministic, that is, running the simulator for
the same $x$ twice will yield the same $y$.

\subsection{Previous 3D capacitive work}
3D pose systems have previously been developed, based on IR and vision systems, e.g. the Leap motion and Kinect devices. \cite{Grosse-puppendahl2017} give a broad overview, and excellent literature survey of capacitive sensing in HCI. An early paper inferring pose with a Bayesian framework is \cite{Smi96}. For mobile devices, capacitive screens have been used. Previous approaches to 3D touch with capacitive sensing in the HCI community included 3D position inference \cite{Rogers2010} and finger pose inference \cite{Rogers2011a}. These used custom hardware and were also the first public demonstrations of 3D pose inference on a standard commercial smartphone at the time (the Nokia N9).  \cite{LeGTayIza14} used random forests to predict pose. The {\it Swiss-Cheese Extended} model proposed in \cite{GroBraKam13} is a real-time method for recognizing objects with capacitive proximity sensors and is demonstrated for the  detection of the 3D-position of multiple human hands in different configurations above a surface that is equipped with a small number of sensors. 

\cite{Hinckley2016} demonstrate `pre-touch', above-surface interaction in a mobile phone sensing configuration similar to that used in this paper. \cite{Wil16} analyses above-device finger tremor characteristics using capacitive sensing. \cite{XiaSchHar15} used classical machine learning techniques to infer 3D pose. \cite{Mur17} gives the first publication  of convolutional networks for pose and position estimation, also in a 3D context. Subsequent work in \cite{MayLeHen17} used convolutional networks on a conventional Android device to infer finger pose, and was followed by related investigations \cite{LeMayBad17,LeMayHen18,MayLeHen18}.

\section{3D Touch inference task}
\label{ch:touch}
To infer finger pose and position away from the touch surface we need a) a sensing technology which can detect the human hand at a distance from the screen and b) an inference mechanism which can estimate the pose and position given the raw sensor readings. 

\subsection{Sensing technology used}
We use a touch screen of dimensions $6.1  \times 9.7$ cm with a prototype transparent capacitive sensor with an extended depth range of between \SIrange{0}{5}{\cm} from the screen (although accuracy decreases with height) and a resolution of $10 \times 16$ pads and a refresh rate of 120 Hz. We sampled at \SI{60}{\Hz}. It was embedded in a functional mobile phone (7.3 $\times$ 13.7cm size), with an external casing making the prototype slightly deeper than normal modern smartphones. The screen was active during the measurements. It is a self-capacitance (as opposed to the more common mutual capacitance) touch screen, with a checkerboard electrode layout. Underneath the electrode plane there is a solid AC-driven guard plane that prevents capacitive coupling between the electrodes and the body of the device. The electrodes are not grounded, they are in the same AC potential as the guard plane.
\begin{figure}[htb!]
\centerline{\includegraphics[width=0.9\linewidth]{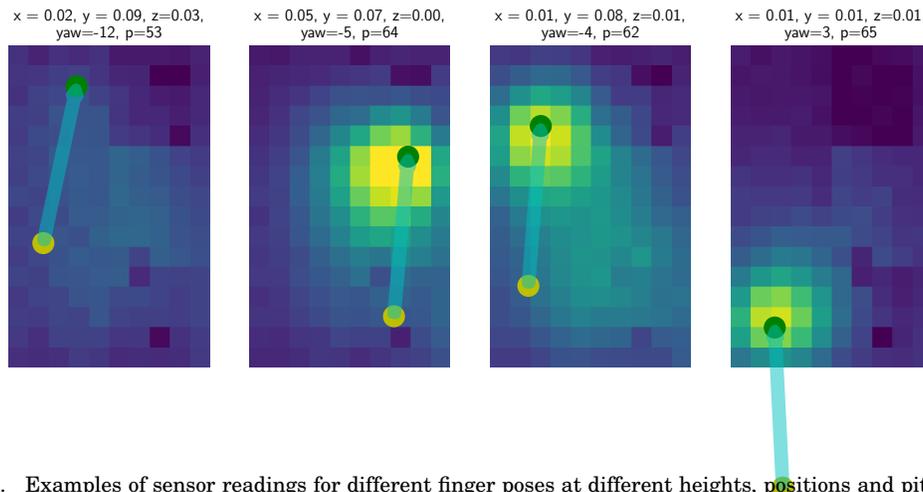}}\vspace{-0.9cm}
\caption{Examples of sensor readings for different finger poses at different heights, positions and pitch and yaw poses. Note how in the 3rd plot the effect of the rest of the hand beyond the finger can be seen in the response.}
\label{fig:illus_poses}
\end{figure}

\subsection{Inference of inverse solutions}
To get a sense of the technical challenge, sensor readings for 4 different finger position and poses are visualised in Figure~\ref{fig:illus_poses}. This paper focusses on the general problem of inference of finger pose in 3D touch from such capacitive sensor data, but the methods could generalise to other sensors. The sensor technologies involved will rarely provide a simple reading which will return the position $(x,y,z)$ and pose (pitch, roll and yaw -- $\theta, \phi, \psi$). Inferring these values can be done with two general approaches:
\begin{enumerate}
\item The creation of a direct complex {\bf nonlinear, multivariable regression mapping}. This is an {\it inverse model} mapping from a possibly high-dimensional sensor-space $X$ to the original $(x,y,z,\theta, \phi, \psi)$ vector.\footnote{In this paper we do not attempt to model the roll angle $\phi$, as that is not feasible with the capacitive technology used in our system.} 

\item The creation of a {\bf causal forward model} mapping from the inputs $(x,y,z,\theta, \psi)$ to the sensor image space $X$, which can then be used to find the most likely values of $(x,y,z,\theta, \psi)$, which would minimise the difference between the observed sensor readings $X$ and the inferred readings $\hat{X}$. If this involves iterative optimisation of a complex forward model, the approach can be computationally intensive.
\end{enumerate}

A significant challenge for methods learning the complex inverse mapping directly from data is that a great deal of training data may be required to learn the mapping, and that many of the approaches used do not predict when they are uncertain about the solution. In the next section, we describe a new approach to solving such inverse problems with reduced demand for training data.

\section{Variational Inference for Inverse Problems}
We use a novel framework to train conditional variational models for solving inverse problems which we introduced in \cite{TonRadTur20}. This approach leverages in combination:
\begin{enumerate}
    \item a minimal amount of experimentally acquired or numerically simulated ground truth target-observation pairs, 
    \item an inexpensive analytical model of the observation process from domain expertise and 
    \item a large number of unobserved target examples, which can often be found in existing data sets. In this way, trained variational inference models can benefit from all accessible useful data, as well as domain models and expertise, rather than relying solely on specifically collected training inputs and outputs. 
\end{enumerate}

\begin{figure*}[t!]
  \centering
  \includegraphics[width=\linewidth]{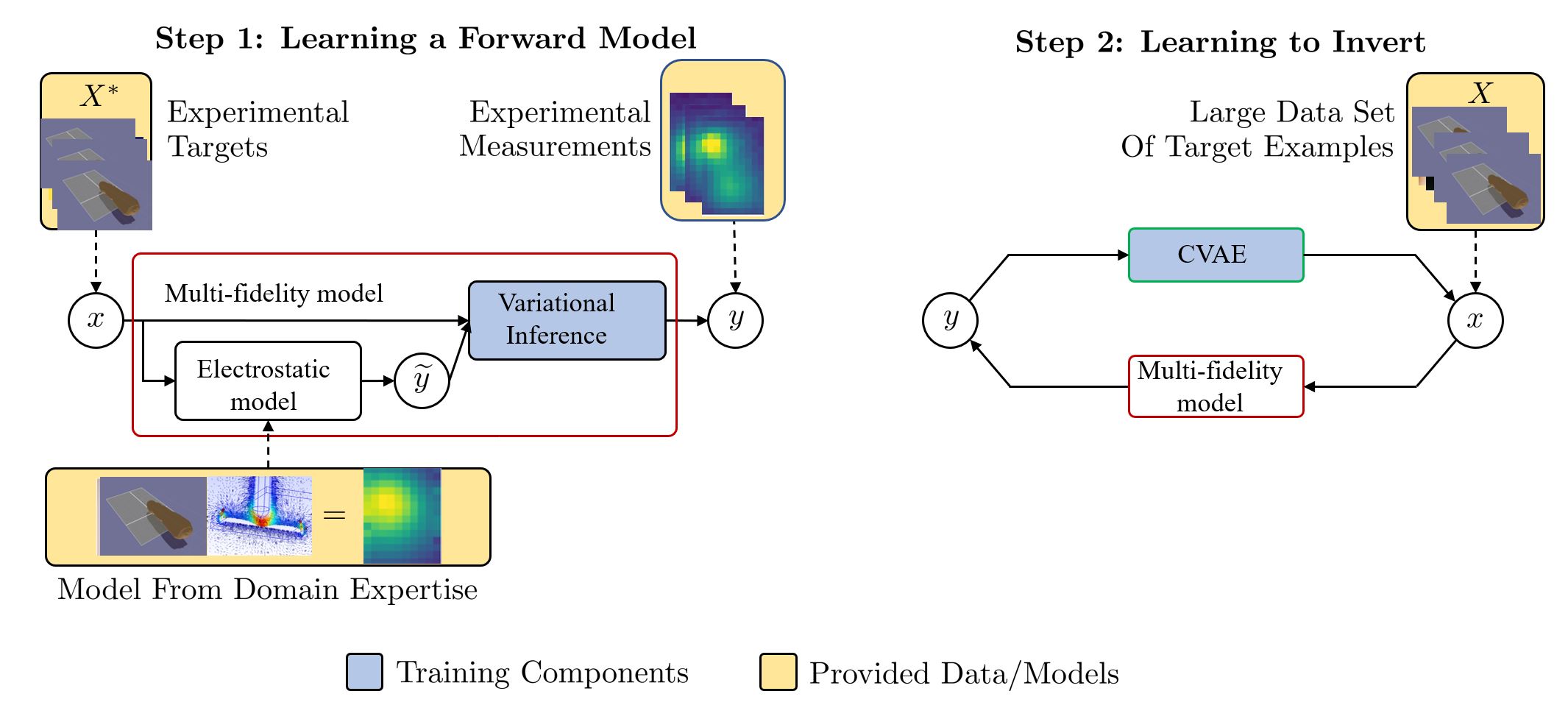}
  \includegraphics[width=0.7\linewidth]{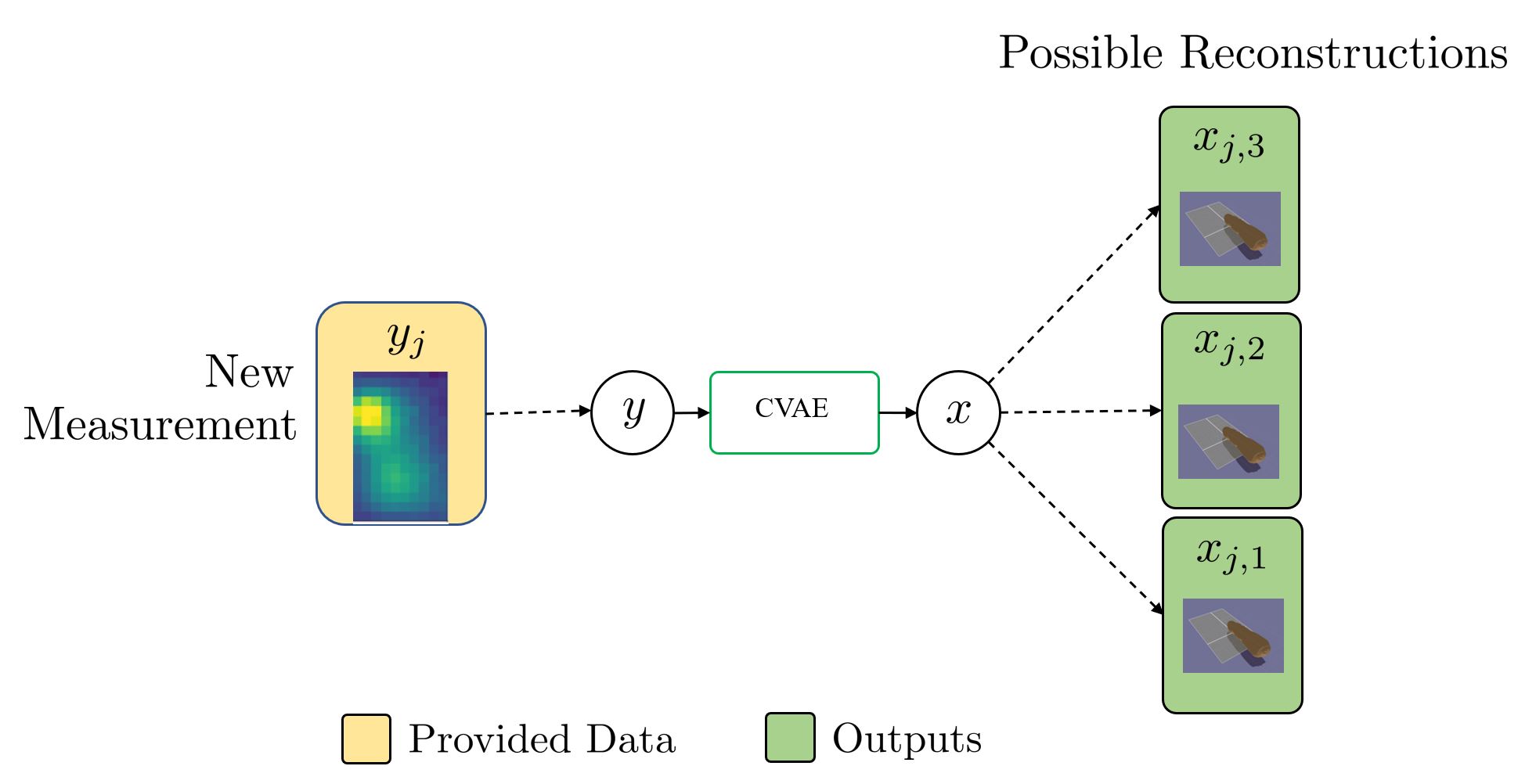}
% \vspace{-0.7cm}
\caption{Proposed framework for training variational inference with diverse sources of information accessible in imaging settings. (a) Firstly, a multi-fidelity forward model is built to generate experimental sensor observations. A variational model is trained to reproduce experimental observations $Y^*$ from experimental ground truth targets $X^*$, exploiting simulated predictions $\widetilde{y}$
 given by some analytical observation model defined with domain expertise. (b) A CVAE then learns to solve the inverse problem from a large data set of target input pose examples $X$ with a training loop; target examples $x$ are passed through the previously learned multi-fidelity forward model to generate simulated sensor measurements $y$, which are then used as conditions for training the CVAE to generate back the targets $x$. In this way, a large number of ground truth targets can be exploited for learning, without the need for associated experimental measurements. (c) The trained CVAE can then be used to draw different possible solutions $x_{j,i}$ to the inverse problem conditioned on a new observation $y_j$.
}
\label{fig:models_2}
\end{figure*}

The framework is derived with a Bayesian formulation, interpreting the different sources of information available as samples from or approximations to the underlying hidden distributions. The training strategy has two stages, followed by a third stage of inference with the trained system:
\begin{enumerate}
    \item Firstly, a {\it`multi-fidelity model' } is used to obtain an estimate of the forward observation process from empirical target-observation pairs and domain models.
Multi-fidelity models are used to predict the high-fidelity outputs of an `expensive' system (where expense can be in computational effort for the model, or experimental effort and cost for data),  from the low-fidelity predicted outputs of an inexpensive, but less accurate system, and their common inputs \cite{MF}.

In this setting, a variational inference model is trained to predict the empirical measurements (high-fidelity outputs) from targets (inputs) and predictions from an observation model built with domain model (low-fidelity outputs). Figure \ref{fig:models_2}(a) illustrates this forward model training stage.
\item Secondly, a {\it conditional variational auto-encoder} (CVAE) is trained to perform the inversion.
CVAEs are latent variable generative models that infer flexible distributions of targets from given conditions, such as labels or image portions \cite{CVAE,PLUG}. In their standard form, they require very large data sets of paired objects and associated conditions to be trained, learning to generate the former from the latter. This is, however, a problem for human--computer interaction tasks, where data acquisition can be time-consuming, and is unappealing to end users.  However, in the framework we propose, training is performed using examples of `target' human inputs alone, as conditions are generated implicitly through the previously learned multi-fidelity model in a loop. 
Given a target input pose, a corresponding sensor observation is computed through the multi-fidelity forward model. The predicted observation is then used as condition for training the CVAE to regenerate the target pose. Figure \ref{fig:models_2}(b) schematically shows the inverse model training stage.
\item The trained CVAE can then non-iteratively draw samples from the probability density function of likely solutions to the inverse problem given a new experimental measurement as input. Figure \ref{fig:models_2}(c) illustrates the inverse inference process.
\end{enumerate}

\subsection{Network architecture}
While the functions used to learn the nonlinear mappings can be from a wide range of architectures, we found both dense multi-layer perceptrons and deep convolutional networks worked well. Deep Convolutional networks (DCN) have a long history \cite{Sch15}, but have made significant progress in recent years \cite{LeCBenHin15,GooBenCou16} through algorithmic improvements and hardware improvements, especially the application of general purpose Graphics Processing Units (GP-GPUs). 2D Convolutional network architectures make the assumption that their inputs are images, allowing us to encode spatial properties into the architecture which constrain it, and make the forward function more efficient to implement and vastly reduce the number of parameters in the network. In this particular case, however, we found better performance with densely-connected networks. These have more representational ability than convnets, and in this task we have (compared to other image processing applications) a relatively low-dimensional image, and there are significant global interactions in the capacitive fields, which affect the inverse problem solutions (edge effects, disturbances from the rest of the hand). 
We therefore used densely connected multi-layer networks with leaky PReLU units, and optimised their architecture (the number of units in each layer) using the ax Bayesian Optimisation framework.

%The grounding question is more interesting because as a mobile device it is not grounded. Instead, there's a voltage difference between the device ground and those electrodes, but the device ground is not necessarily the same as mains ground. Only in the case where the phone is connected to a charger the grounds are the same. So there's typically a voltage difference between the user (typically at mains ground) and the device ground, and an another voltage difference between the user and the sensor electrode, and it is the combined effect of the capacitive coupling between the device ground and the user and the capacitive coupling between the sensor electrodes and the user that is sensed with the electrodes

\section{Generating representative data}

In order to learn the mapping between sensor inputs and finger poses we need a large, carefully calibrated training set of fingers in different poses. For 3D inputs, the mappings can be extremely complex, so generating exhaustive data with human users is effort and time intensive. We therefore explored three approaches:
\begin{enumerate}
\item Robot-generated inputs 
\item Larger sets generated by an electrostatic simulator
\item Human-generated inputs with augmented optical tracking of finger pose and location.
\end{enumerate}

\subsection{Training set 1: Robot data}
The initial explorations were based on robot-generated data. The robot used a 7mm brass cylinder with a hemispherical tip to simulate a finger placed on the screen of the device. A photo of the robot in operation can be seen in Figure~\ref{fig:data} (a). The panel and robot co-ordinates were in different co-ordinate systems. We tested the $(x,y)$ returned by the current prototype hardware API as an initial stage of process verification. The resulting plots of the robot and panel co-ordinates can be seen in Figure~\ref{fig:data} (b), (c).
\begin{figure}[htb!]	
	\subfigure[The robot configuration]{
		\includegraphics[width=0.63\linewidth]{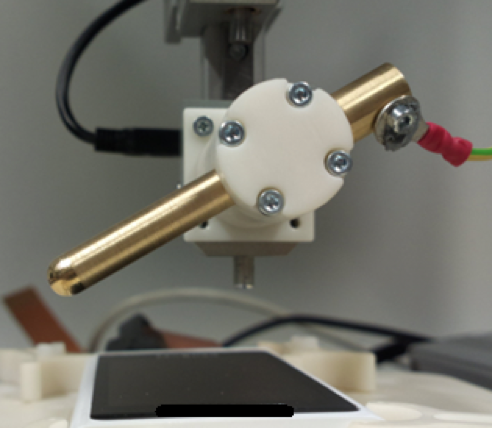}
	}
	\begin{minipage}[b]{0.28\linewidth}
		\subfigure[$x$ data]{	
			\includegraphics[width=\linewidth]{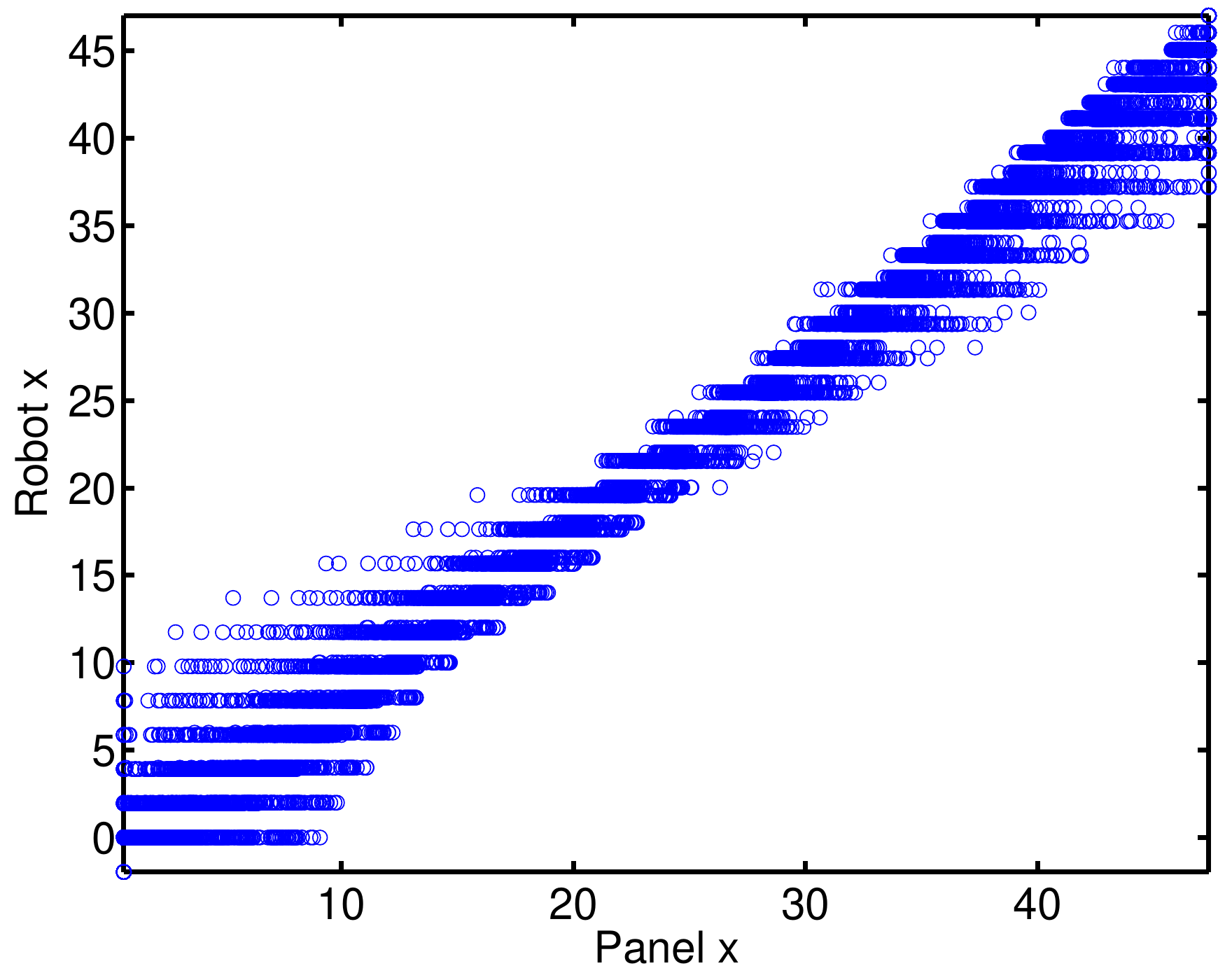} 
		}
		\subfigure[$y$ data]{
			\includegraphics[width=\linewidth]{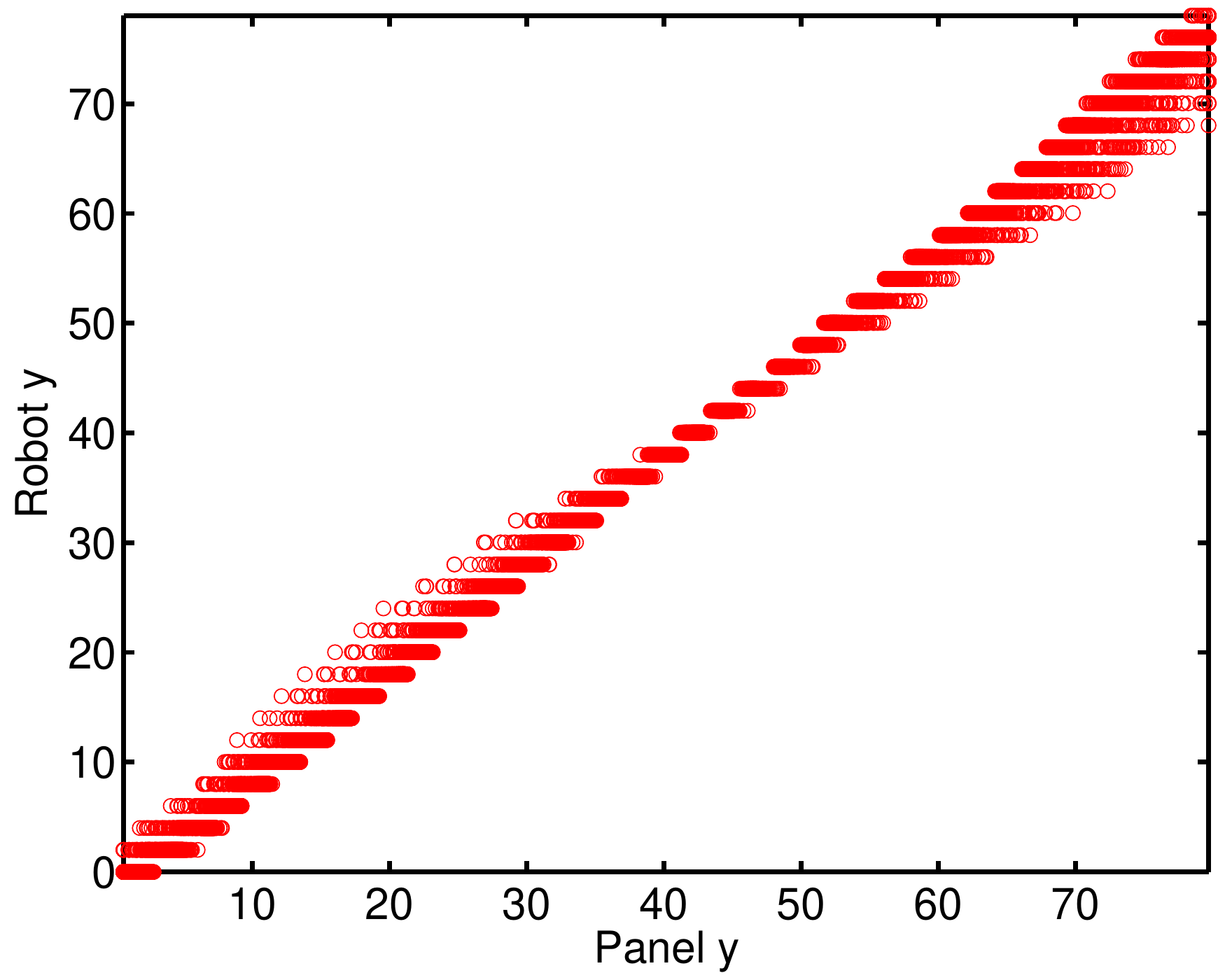}				
		}
	\end{minipage}		
	\centering\caption{\label{fig:data}Robot versus touch panel API co-ordinates.\label{fig:data}}
\end{figure}

%Both the sensor data and target data were pre-processed. The sensor data was processed such that each sensor had mean zero and standard deviation one. Values for $x,y,z,\theta_p$ were pre-processed to lie in the range $[-1,1]$ ($\theta_p$ is the pitch angle).

The use of robot-controlled inputs has the advantage of potentially allowing repeatability, and precise control of the position and pose of the pointer. It also uses the real sensors for a given device. 

The use of the final sensors can be both an advantage and a disadvantage. It means that the noise distributions are more realistic, but it also requires the final sensor to be complete before analysis of expected algorithm performance and robustness can be made. With our particular setup, we found the creation of training sets using robots to be a costly and time-consuming process, which often had errors, and which became a limitation to exploratory research, while still having limitations in terms of realism for the target use case of inferring human finger pose. Because of this we developed a parallel approach based on testing the suitability of data produced by electrostatic simulators. 
%\FloatBarrier
%\section{Electrostatic models}

\subsection{Training set 2: Electrostatic simulation}
Simulations will inevitably make approximations in describing the pointing objects and the physics of the system. They are, however, very useful for testing possible configurations, and predicting how many observations are needed to calibrate a new system, or to customise a device to a specific user. 

The simulation was performed using a 3D finger object created by attaching a hemisphere to one end of a cylinder. The default diameters of the cylinder and hemisphere were both \SI{9}{\mm}, while the total length of the finger object was \SI{10}{\cm} (although all of these dimensions were defined as script parameters and could be easily changed). This script runs finite element method (FEM) simulation and parses the meshed charge values into total charge per plate values. These values represent the charge/capacitance matrix. 
%\begin{figure}[tbh]
%	\includegraphics[width=0.5\linewidth]{layout_topdown.pdf}			%\centering\caption{\label{fig:layout_top} Top-down view of the simulator, showing how yaw angles are defined.}
%\end{figure}
%\begin{figure}[tbh]
%	\includegraphics[width=\linewidth]{layout_side.pdf}			%\centering\caption{\label{fig:layout_side} Side-on view of the simulator, showing how pitch angles are defined.}
%\end{figure}
%Figures~\ref{fig:layout_top} and~\ref{fig:layout_side} show the simulator as it appears while processing a selected point. Figure~\ref{fig:layout_top} gives a top-down view, with 
The $(x, y)$ origin is defined as the bottom left corner of the ``screen'' and the yaw $\psi$ is defined as the angle the ``finger'' makes with the $y$-axis, with \ang{0} straight up the $y$-axis, negative yaw rotate to the left, and positive yaw values would rotate the finger towards the right. %Figure~\ref{fig:layout_side} shows a similar finger position viewed from the side, looking down the $x$-axis of the screen. 
The pitch $\theta$ is defined to be \ang{0} when the finger is positioned vertically, perpendicular to the screen, and \ang{90} when parallel to the screen. The figure shows a pitch of around \ang{60}. Note that with this representation of a finger, when pitch=\ang{0}, all yaw angles are indistinguishable. For our purposes we wished to be able to position the lowest point of the ``fingertip'' over the given $(x, y, z)$ position, as opposed to other possibilities such as centre of the ``fingertip''.

To obtain simulated measurement data, we solve for the following electrostatic boundary value problem that models the exterior domain of the mobile phone. The phone is modelled as two parallel equipotential planes. The upper plane represents the sensor electrode layer beneath the screen surface, kept at a constant voltage by the phone electronics. The lower plane represents the rest of the phone mechanics at a ground reference voltage. A human finger interacting with the device is modelled as a cylinder with half-spheres at the ends. The finger surface is set to the same reference voltage as the phone ground. We control both the location of the fingertip as well as the finger angles to simulate different finger postures.

\begin{figure}[htb]
\centering
\includegraphics[width=0.8\linewidth]{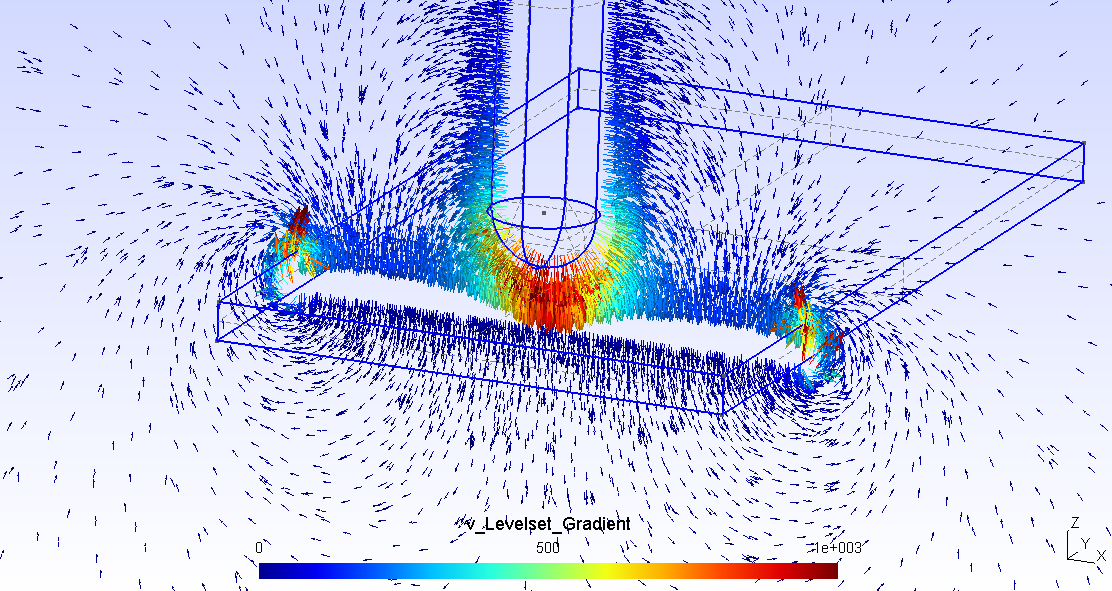}
\caption{Field plot from electrostatic simulation}
\end{figure}
\begin{figure}[htb]
\centerline{\includegraphics[width=0.8\linewidth]{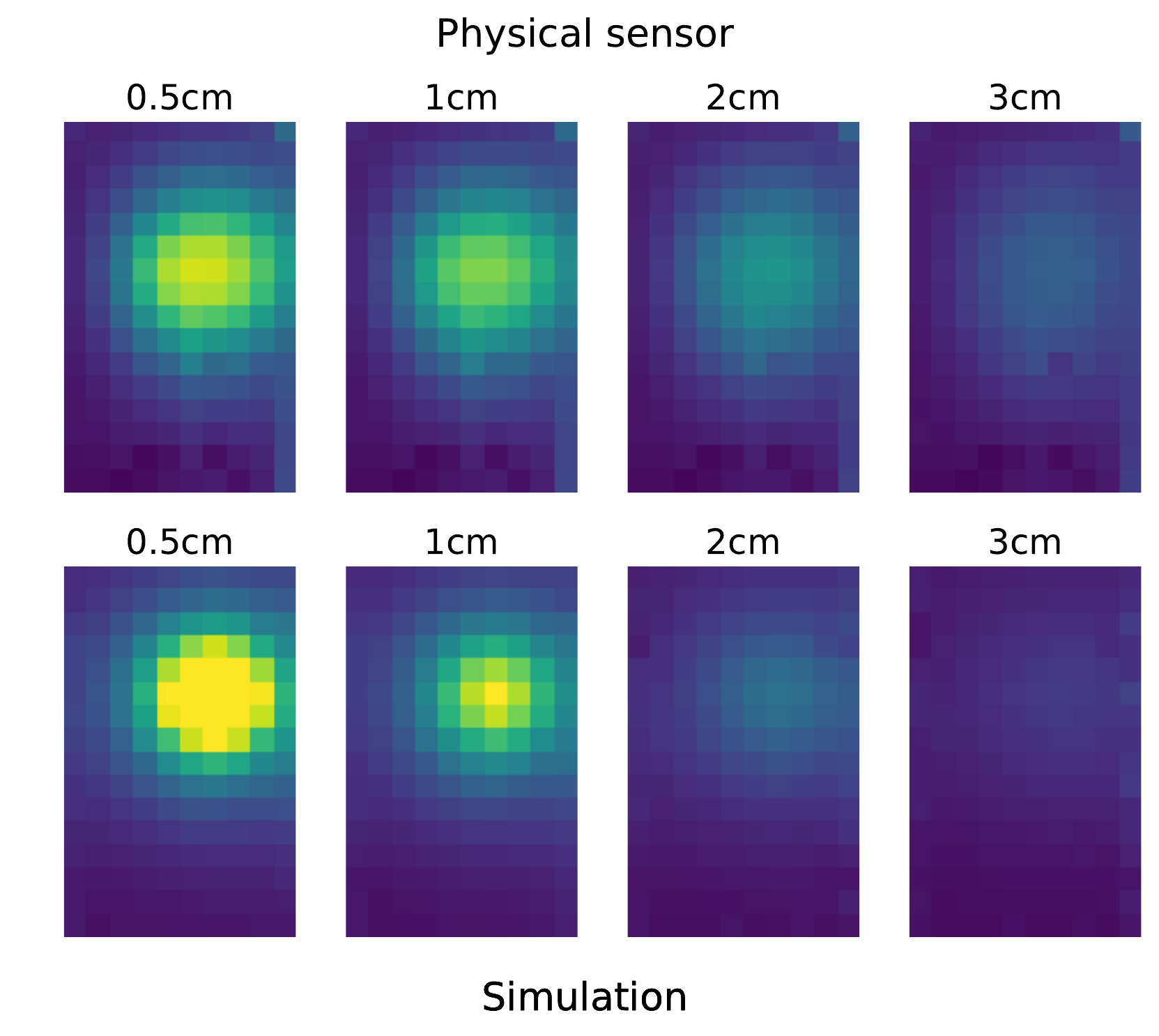}}
\caption{Real sensor data from the device, and Simulated sensor data for comparison, in the range $z$=0.5-3cm. }
\label{fig:sensor_sim_images}
\end{figure}

\begin{figure}[htb]
\centerline{\includegraphics[width=0.8\columnwidth]{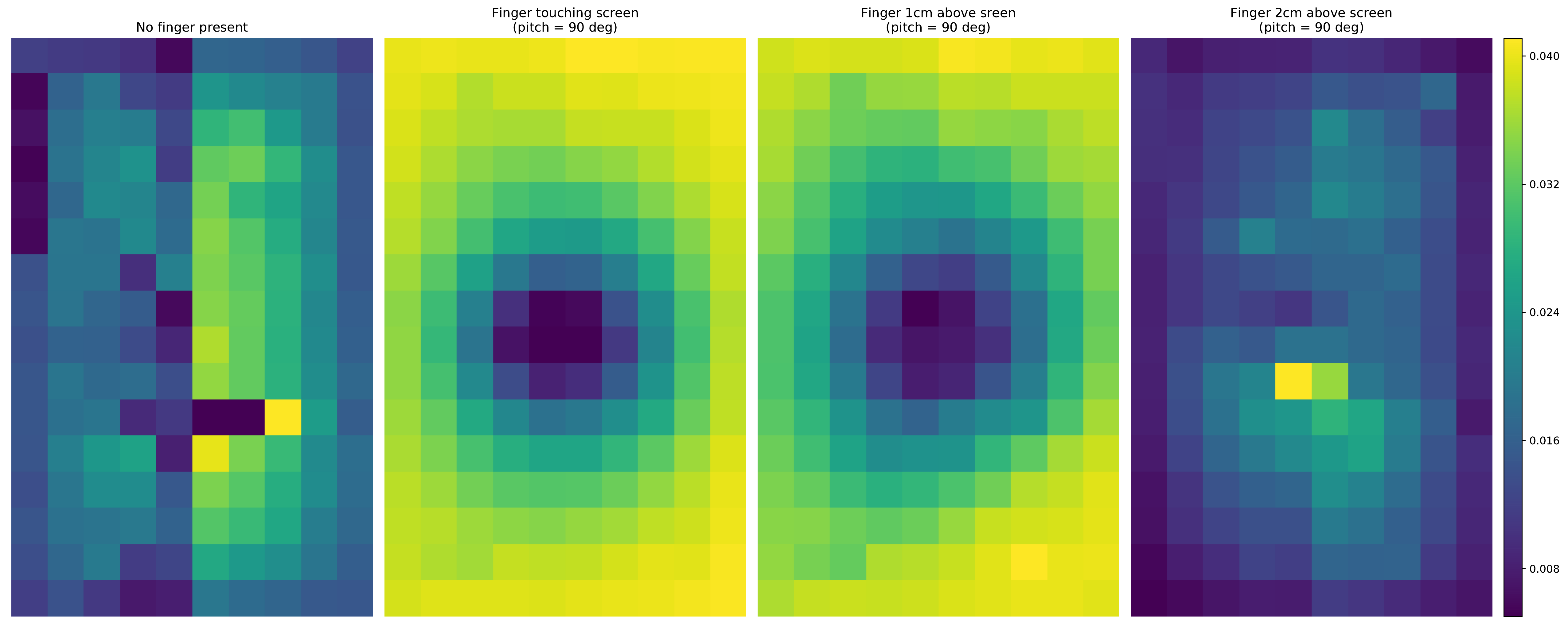}}
\caption{A map of standard deviations of pixels on the physical sensor with, from left to right: no finger present, vertical finger on the screen, vertical finger 1cm above screen, vertical finger 2cm above screen (for sensor data normed to the 0-1 range).}
\label{fig:noisemap}
\end{figure}

 
Such an arrangement generates an electric field in the phone exterior domain, and the capacitive sensor measures the net charge that is induced on each individual pad by the electric field. In the electrostatic model, the net charge is computed as the integral of the electric flux across a pad surface. We use the finite element method to solve for the electrostatic problem, and sample the electric flux on each pad to approximate the sensor readings. The simulation used was provided by our industrial partners, involving open source electrostatic 
simulation models (Gmsh \cite{GeuRem09} and GetDP\footnote{\url{http://www.geuz.org/getdp/}}).  While the model is a crude approximation of the actual phone and a finger geometry,  the error in the the simulated sensor readings are on par with the noise level in the actual measurements from prototype devices, as shown in Figure \ref{fig:sensor_sim_images}. This gives us some insight into the differences between the simulated and sensed data. These simulations used the settings of a \SI{10}{\cm} finger with a width of \SI{7}{\mm}. The simulation is most accurate between \SIrange{1}{3}{\cm} -- beyond that and we begin to see what appear to be errors related to numerical issues.
Similarly these results do not simulate the sensor noise. The physical observations in Figure~\ref{fig:sensor_sim_images} are smoother than real individual frames, as the values were averaged over 6 seconds, from 3$\times$2 second capture sessions with the finger held stationary. A sample of noise on the real sensor was recorded, allowing us to show the standard deviation in Figure~\ref{fig:noisemap}. The noise on individual pixels is not correlated with others, and generally had a positive mean offset, and Gaussian distribution. The standard deviation is smallest at the edges of the screen. Note, however, how the noise changes as the finger touches, then moves away from the screen (being held static in each pose for 30s). The pixels closest to the finger show less variability, while those around them show increased variability.

%\begin{figure}[htb]
%\centering
%\subfigure[Real sensor data from the device.]%%{\label{fig:device1}
%\includegraphics[width=0.45\linewidth]{device_12mm.png}}
%\subfigure[Real sensor data from the device]%{\label{fig:device2} 
%\includegraphics[width=0.45\linewidth]{device_01mm.png}}
%caption{Data from the real sensor. The fingertip 
%position is marked by the green circle, and the purple line shows the position of the finger itself.}
%\end{figure}

\subsection{Training set 3: Human-generated input}
\begin{figure}
\centerline{\includegraphics[width=0.8\linewidth]{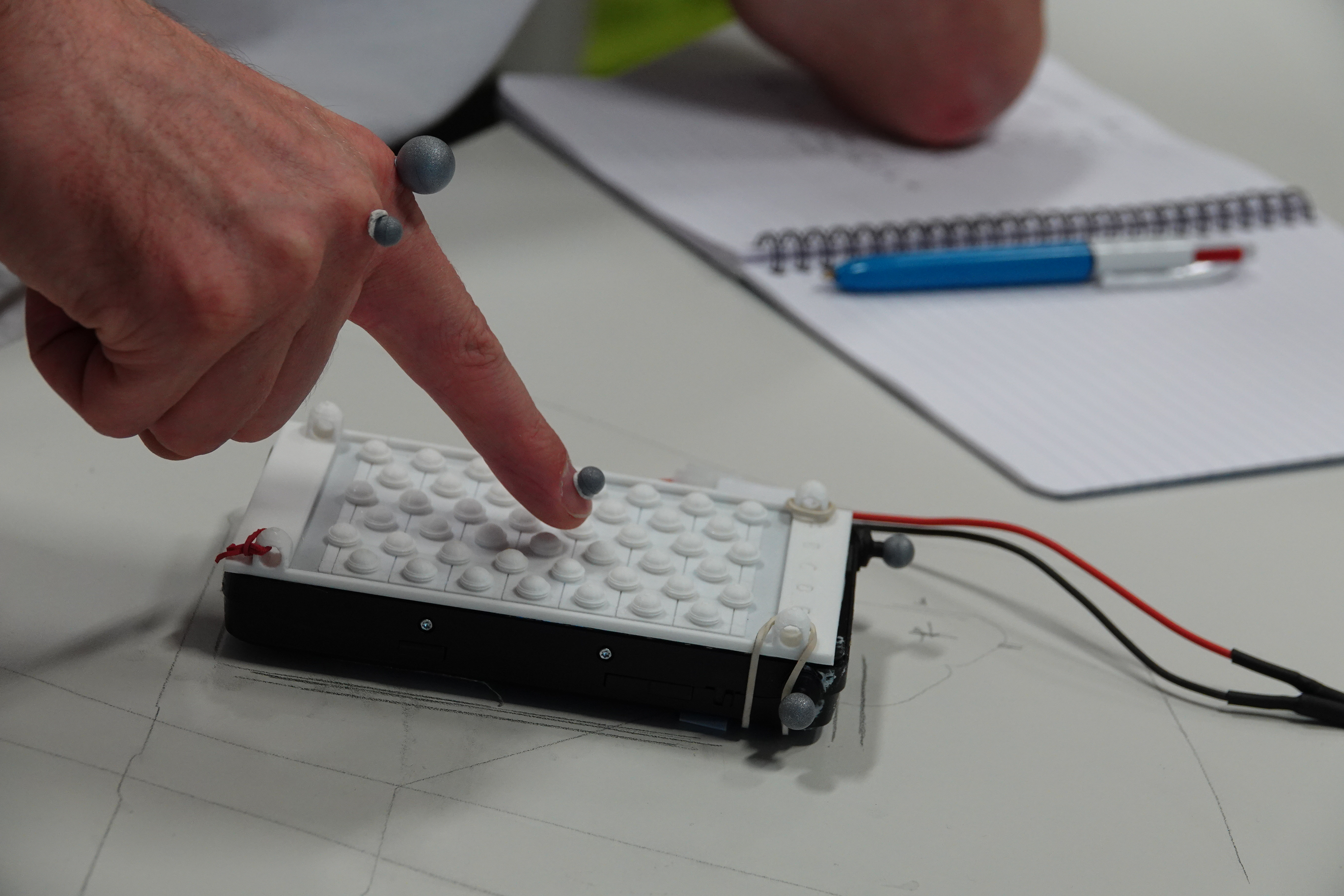}}
\caption{Lab setup -- human input with optical tracking. 3D printed calibration cover was removed for final trials.}
\label{fig:humaninput}
\end{figure}

Creating comparable datasets with human intervention will provide the most realistic sensor data, but it also requires an alternative technology to sense finger pose and position, and it will be time-consuming and difficult to obtain accurate poses in all the desired configurations, and there is still significant potential for errors and sensor noise to enter the analysis. In the same style as used in \cite{MayLeHen17,MayLeHen18}, we use an Optitrak system to accurately track the hand pose and position, as shown in Figure~\ref{fig:humaninput}.  

\newpage
\section{Empirical modelling results}

\subsection{Emulation of Simulation model}
Our first application of machine learning is to create an emulator of the physics-based model which is able to run with reduced computational expense and hence higher speed. We therefore generated some simulated data to train the emulator model. 

\subsubsection{Simulated training set structure}
The training data was generated as follows: 15973 points randomly distributed in a rectangle with bounding box between $(x,y,z) = (1.0, 2.0, 0.3)$ and $(5.0, 10.0, 3.0)$ cm,  pitch angles are distributed between \ang{0}--\ang{90} and yaw between \ang{-90}to \ang{+90}. In addition, 25830 points at positions and heights every 0.5cm in the range $0.05$--$5$cm in $x$, $1$--$8$cm in $y$ and every 0.5cm from $0.3$--$3.5$cm in $z$. At each of these positions, all combinations of pitch angles 
$[\ang{0}, \ang{15}, \ang{30}, \ang{45}, \ang{60}, \ang{75}, \ang{90}]$, and yaw angles 
$[ \ang{-45},\ang{-30}, \ang{-20}, \ang{-10}, \ang{0}, \ang{10}, \ang{20}, \ang{30}, \ang{45}]$ were tested. This led to a total training set of 41803 simulated points. In addition, simulations were generated for all human-generated points (see \ref{sec:humangen}) which created a further 40428 training, and 41237 test simulation points.

%560 points were excluded from the inverse, or regression training set for pitch angles of \ang{0} from vertical with non-zero yaw angles, as these are indistinguishable. These are kept in the forward model (to help ensure that model accurately represents the yaw invariance, to avoid pointless adaptation of yaw in the vertical condition). Similarly data beyond the 5cm limit, or for roll angles (which cannot be differentiated by the inverse model) can be included in the forward model. #CHECK I AM STILL REMOVING THESE FROM THE INVERSE TASK

%, as shown in Figure \ref{fig:training_data}, which we typically split 80/20 into training and test sets.
%To experiment with the central area of the screen (where there is more information for yaw), we also created a subset, {\bf `Central set'} of  * points found in the bounding box between (2,2,0.3) and (4,4,3).
%\begin{figure}[htb!]
%\vspace{-0.3cm}
%\includegraphics[width=\linewidth]{training_data.pdf}
%\vspace{-0.7cm}
%\caption{Training data as quiver plot. This gives an impression of the combination of a systematic grid of positions and angles, and random position and pose points.}
%\label{fig:training_data}
%\end{figure}

\subsubsection{Network architecture}
The network was implemented in Python, using Tensorflow \cite{tensorflow2015-whitepaper}. This is a flexible library, which allows efficient use of GPU acceleration. 
As described earlier, we use the same CVAE architecture described in \cite{TonRadTur20}, with the numbers of units in each layer optimised with a Bayesian Optimisation approach, using the ax\footnote{\url{https://ax.dev/docs/bayesopt.html}} framework. %This led to an architecture with [ DETAILS].

\subsubsection{Simulated data modelling results}
\begin{figure}[htb]
\includegraphics[width=\linewidth]{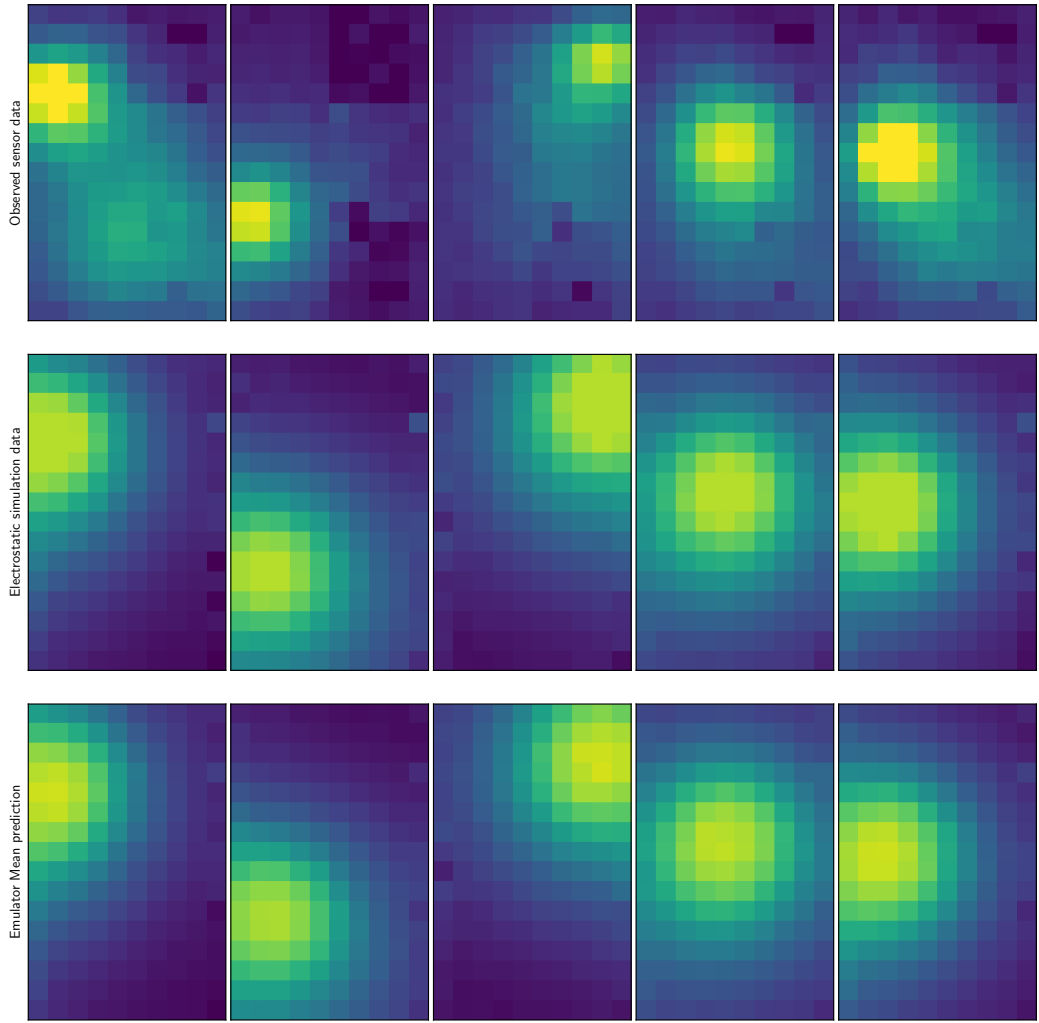}%{real_sim_predicted.pdf}
\caption{Forward simulation emulation. Comparison of real sensor data (upper row), with electrostatic simulation outputs (middle row) and the deep network emulation model of the simulation for the same pose (lower row). The red marker on the upper row indicates the fingertip position.}
\label{fig:compareSim}
\end{figure}

The RMSE on test sets of the overall data was 0.013. Compare the simulation model predictions with the network responses, in Figure~\ref{fig:compareSim}, showing a good match between the first-principles simulation results and the black-box model predictions. It also did this with a speedup of 2.4 million, completing predictions for 20410 inputs in 0.67s on the same machine that took 80s per prediction for the simulation.

\subsection{Learning a forward model}
We now move to use human-generated data, together with the emulator of our electrostatic simulation to create a multi-fidelity forward model of the mapping from a human pose to the predicted sensor outputs.

%\subsubsection{Robot data}
%\subsubsection*{Robot training set structure}
%We acquired 20,000 training examples from a $8 \times 13$ sensor array, on a 2mm grid size. At each grid location, the sensor data was recorded with the simulated finger at a series of heights: 0, 1, 2, 3, 5, 10, 20 and 30mm (measured from the surface of the screen). This whole process was performed once each for a series of different finger angles: \ang{0}, \ang{5}, \ang{15}, \ang{30}, \ang{45} and \ang{60} (measured from vertical).

%\subsubsection*{Robot data modelling results}

%\begin{itemize}
%    \item No simulation results RMSE
%    \itemWith simulation data RMSE
%\end{itemize}

%\begin{figure}
%    \centering
%   \caption{Plot of change in error for the forward model of the robot-generated data, with increasing amounts of training data for the cases of simulated data and no simulated data.}
%    \label{fig:transferRobot}
%\end{figure}

%\begin{figure}
%    \centering
%%    \includegraphics{}
  %  \caption{Examples of responses }
%    \label{fig:my_label}
%\end{figure}

%\begin{lstlisting}
%x = Dense(160,activation='relu')(input_vec)
%x = Dense(160,activation='relu')(x)
%x = Reshape((1, 10, 16))(x)
%x = Convolution2D(10, 4, 4, activation='relu', border_mode='same')(x)
%x = Convolution2D(16, 3, 3, activation='relu', border_mode='same')(x)
%x = Convolution2D(16, 3, 3, activation='relu', border_mode='same')(x)
%x = Convolution2D(16, 3, 3, activation='relu', border_mode='same')(x)
%decoded = Convolution2D(1, 3, 3, activation='linear', border_mode='same')(x)
%\end{lstlisting}
\subsubsection{Human-generated data}
\label{sec:humangen}
We asked users to systematically perform a variety of tasks, as enumerated in Table~\ref{tab:my_label}, which were intended to be easy to explain and cover a range of realistic inputs. We logged 18 sessions, where the task was repeated twice in separate sessions. Tasks 6 \& 7 and 8 \& 9 are also identical. We then split these sessions into test and training sets, so that they both contained the type of movement, but there was no overlap in actual movements.
\begin{table}[htb!]
    \tbl{Tasks in the human trials}{
    \label{tab:my_label}
    \centering
    \begin{tabular}{l|l}
    {\bf Task \#} & {\bf Task Description}\\
         0,1 & Touching all 60 points on the paper overlay with minimal yaw\\
        2,3 & Touching all 60 points on the paper overlay with negative yaw (finger coming in from left)\\
        4,5 & Touching all 60 points on the paper overlay with positive yaw (finger coming in from right)\\
        6,7 & Randomly moving fingertip over device (with occasional touches) for 60s\\
        8,9 & Randomly moving fingertip over device (with occasional touches) for 60s\\
        10,11 & Touching corners of screen, moving along edges and diagonals\\
        12,13 & Touch 7-8 random points with some yawing while touching (positive and negative)\\
        14,15 & Keep finger touching screen, and move through (close to) horizontal--vertical range,\\
        & with minimal yaw\\
        16,17 & Typing-style random touches all over the screen for 60s \\
    \end{tabular}
}
\end{table}

The test users systematically performed up/down button tapping motions at a grid of $x,y$ positions and in a range of pitch and yaw angles, with their finger extended, as shown in Figure~\ref{fig:humaninput}. This requires further signal processing to transform the positional data of the markers into robust estimates of the tip of the finger, and the pitch and yaw angles.

\subsubsection*{Distribution of the data}
\begin{figure}[htb!]
    \centering
    \includegraphics[width=\linewidth]{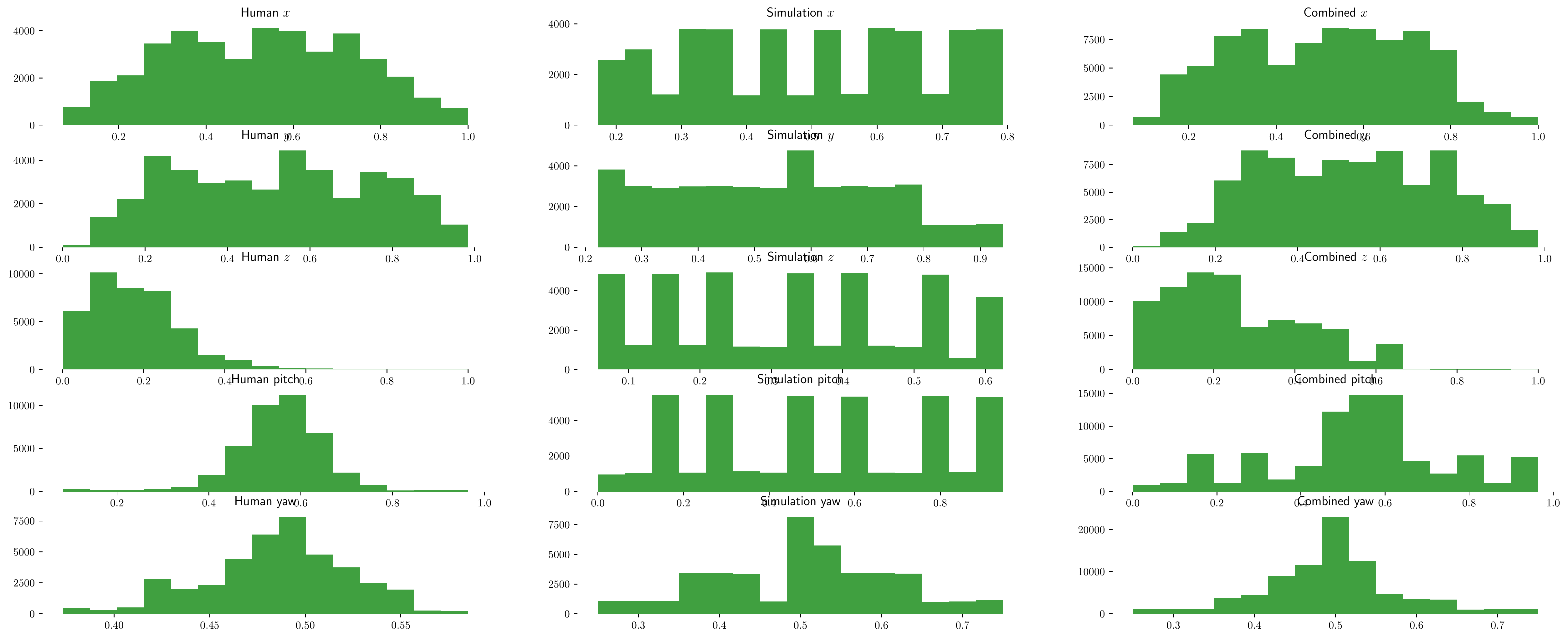}
    \caption{Histograms of the different variables in human-generated data, simulation-only and combined datasets.}
    \label{fig:histdata}
\end{figure}

With such a flexible input sensing mechanism it is difficult to guarantee that all relevant poses have been observed, without a very time-consuming data acquisition process, which would also include a lot of control of the human subject's motions, which has the potential to be counterproductive in some cases. 
Figure~\ref{fig:histdata} shows histograms of the different inferred states in human-generated and simulation+human generated. Although we had encouraged the test users to vary pitch and yaw throughout, we can see that the sampled data is unevenly distributed, with regions of yaw space sparsely covered. This highlights the potential benefits of using simulation together with real data, as the simulated data can be more widely spread around the space, with the potential to make the system more robust.

\subsubsection{Network architecture}
We used the same architecture as for the simulated model above, again with numbers of units in each layer optimised with a Bayesian Optimisation approach.  Adam optimisation was used, with batch size = 32. %This led to an architecture with: $n_z=5$, for the $x$ encoder, layers of 22, 40, 45 =, for the low-fidelty encoder, layers with  84, 81, 61, for the encoder, layers with 83, 65, 46, 12, for the high-fidelity encoder  116, 88, 46, for the low-fidelity decoder 195, 136, 39, for the latent space decoder, layers with 43, 31, 19, 'N_dec1': 91, 116, 165, 'N_x_dec1': 13, 54, 73, and an initial training rate = 1e-05,  and batch size = 128.

%We used the same training data locations as in the previous section, with the addition of multiple yaw values at the \ang{0} case.  

%$\ell_2$ weight penalisation and batch-normalisation were used, but other regularisation techniques, e.g. Dropout were found not to improve performance or robustness. [UPDATE]

\subsubsection{Human-generated data modelling results}

\begin{figure}[htb]
\centerline{\includegraphics[width=0.8\linewidth]{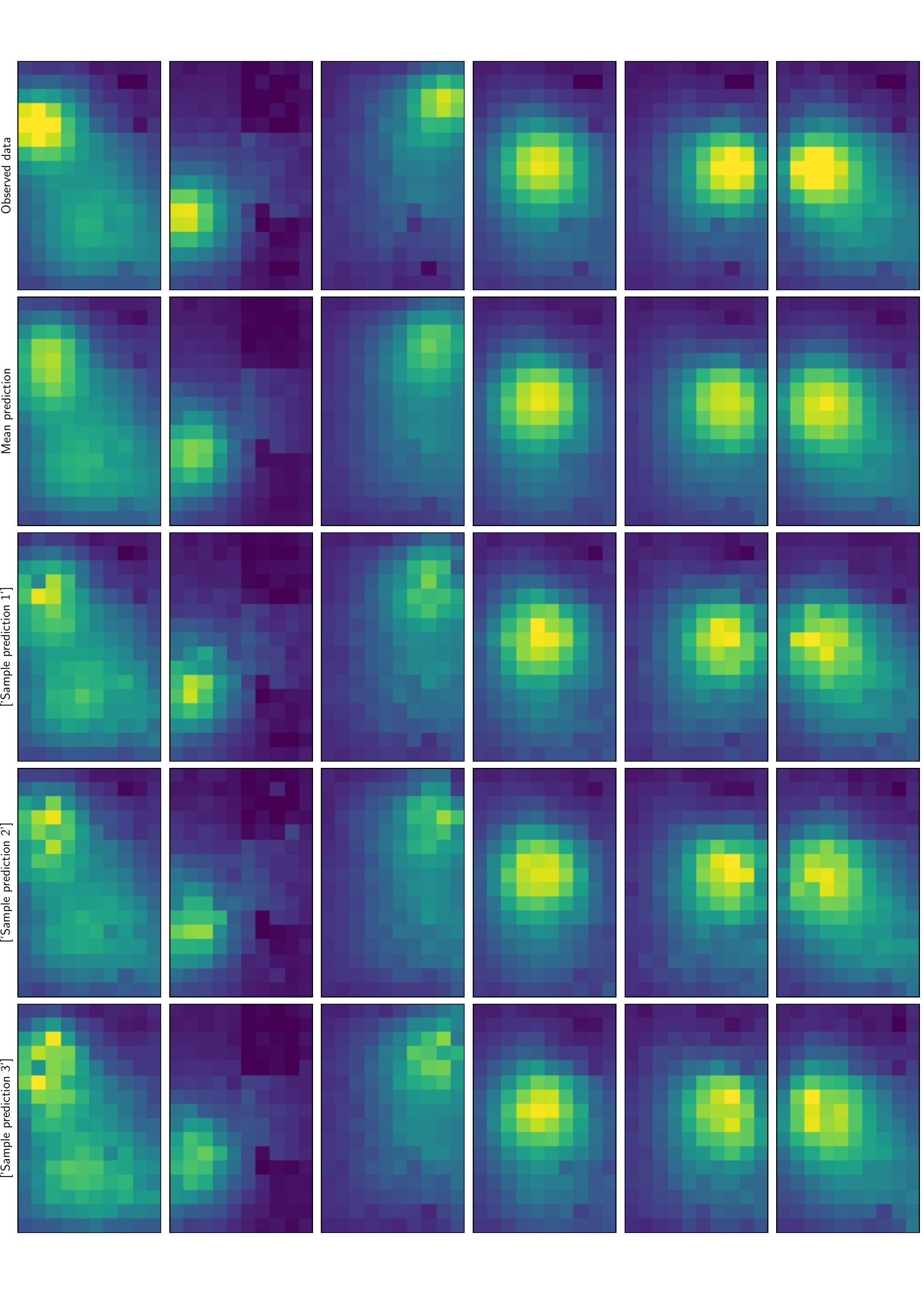}}%\vspace{-2cm}
\caption{VAE Forward Model. Comparison of sensed outputs (top row) and the learned forward model of the same pose by the human. The second row is the mean response from the forward model, and later rows in each column are examples of random samples from the model for the same input pose. Note that these capacitive responses are significantly more complex than traditional touch screen touch responses. The model has learned to correctly reproduce faulty, nonresponsive pixels in the prototype hardware.}
\label{fig:comparehumint}
\end{figure}

The results on human generated data are shown in Figure~\ref{fig:comparehumint}. These show how well the networks can predict the expected sensor image from a given finger pose, despite the responses being significantly more complex than typical capacitive screens exhibit, when a single finger is involved. Note that the model learns to correctly reproduce things like the dead pixels in the prototype hardware at bottom and top right corners of the images. The mean images are very close to the target images, although often less intense at the area of peak intensity. The individual samples are slightly overly noisy and less representative of the observed data, indicating scope to further refine the forward model, possibly with more data or a different model architecture. 

\begin{figure}
    \centering
    \centerline{\includegraphics[width=0.9\linewidth]{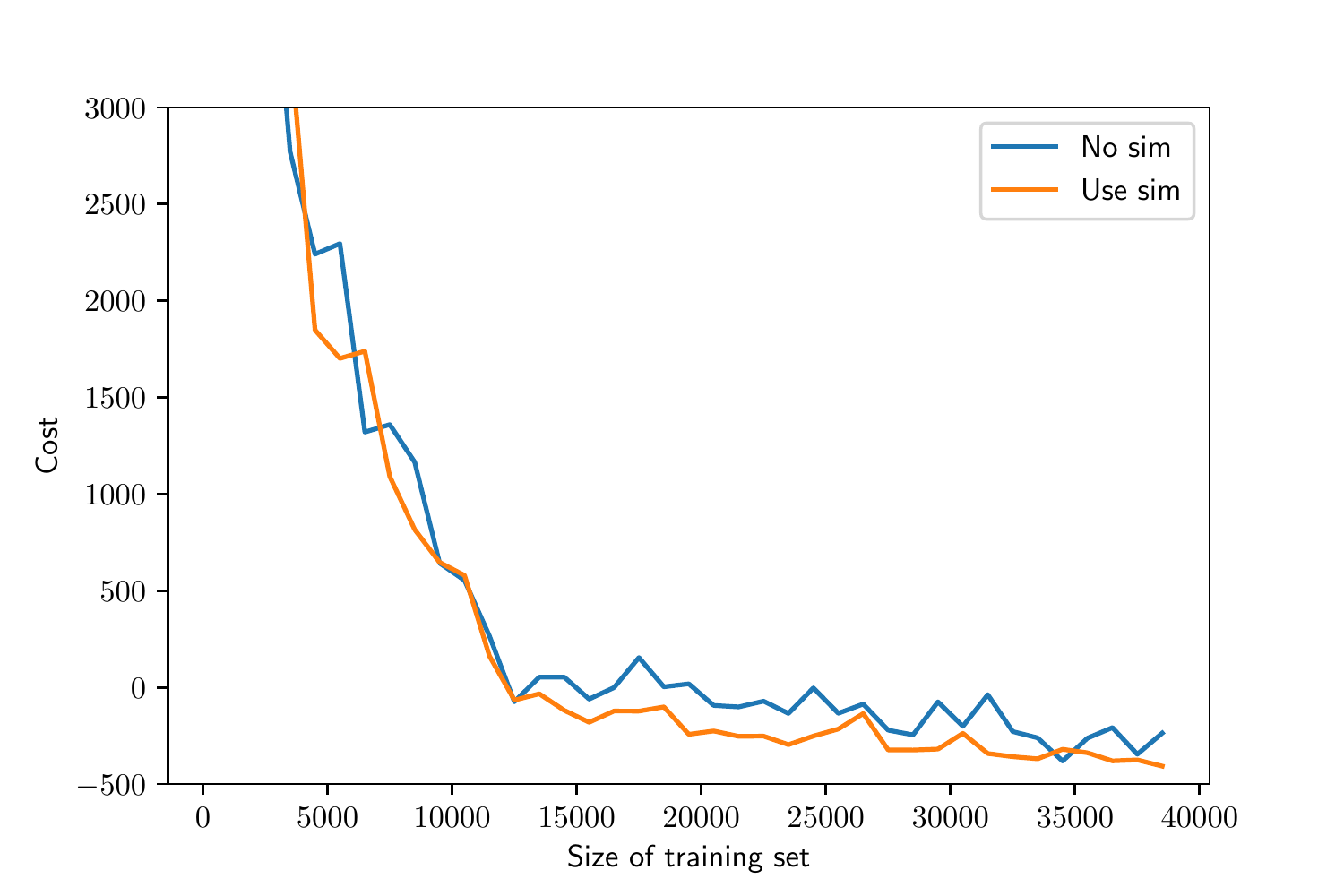}}
    \caption{Plot of change in cost (negative Evidence Lower Bound, ELBO) on a test set of data for the forward model of the human-generated data, with increasing amounts of training data for the cases of use of an additional low-fidelity simulator and no simulator. For smaller training sets the additional simulator input has no clear advantage, but as the model improves, from ca 10,000 onwards, its use brings a clear benefit.}
    \label{fig:transferHuman}
\end{figure}

\begin{figure}
    \centering
    \includegraphics[width=0.95\linewidth]{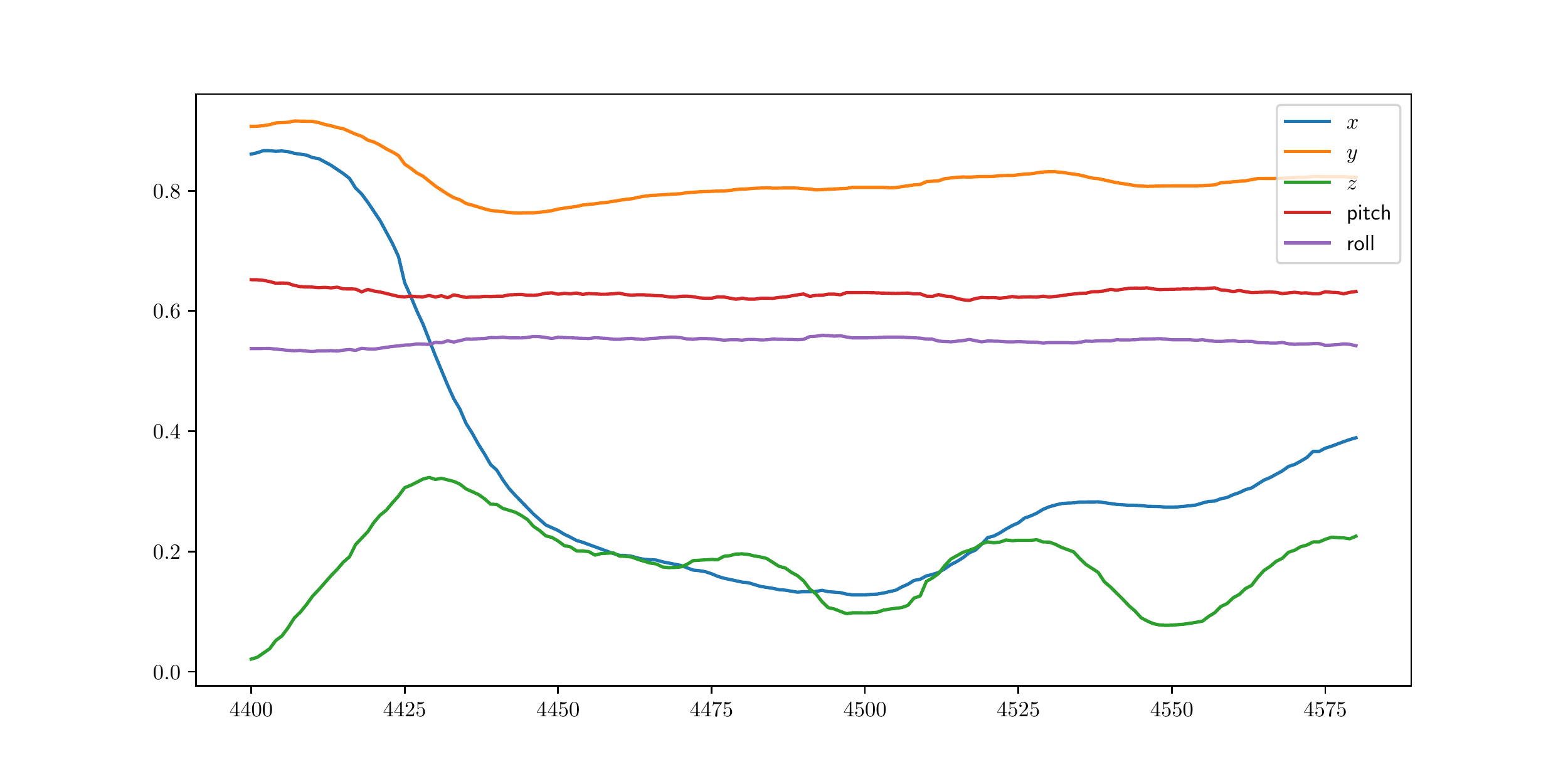}\vspace{-.4cm}
   \includegraphics[width=0.8\linewidth]{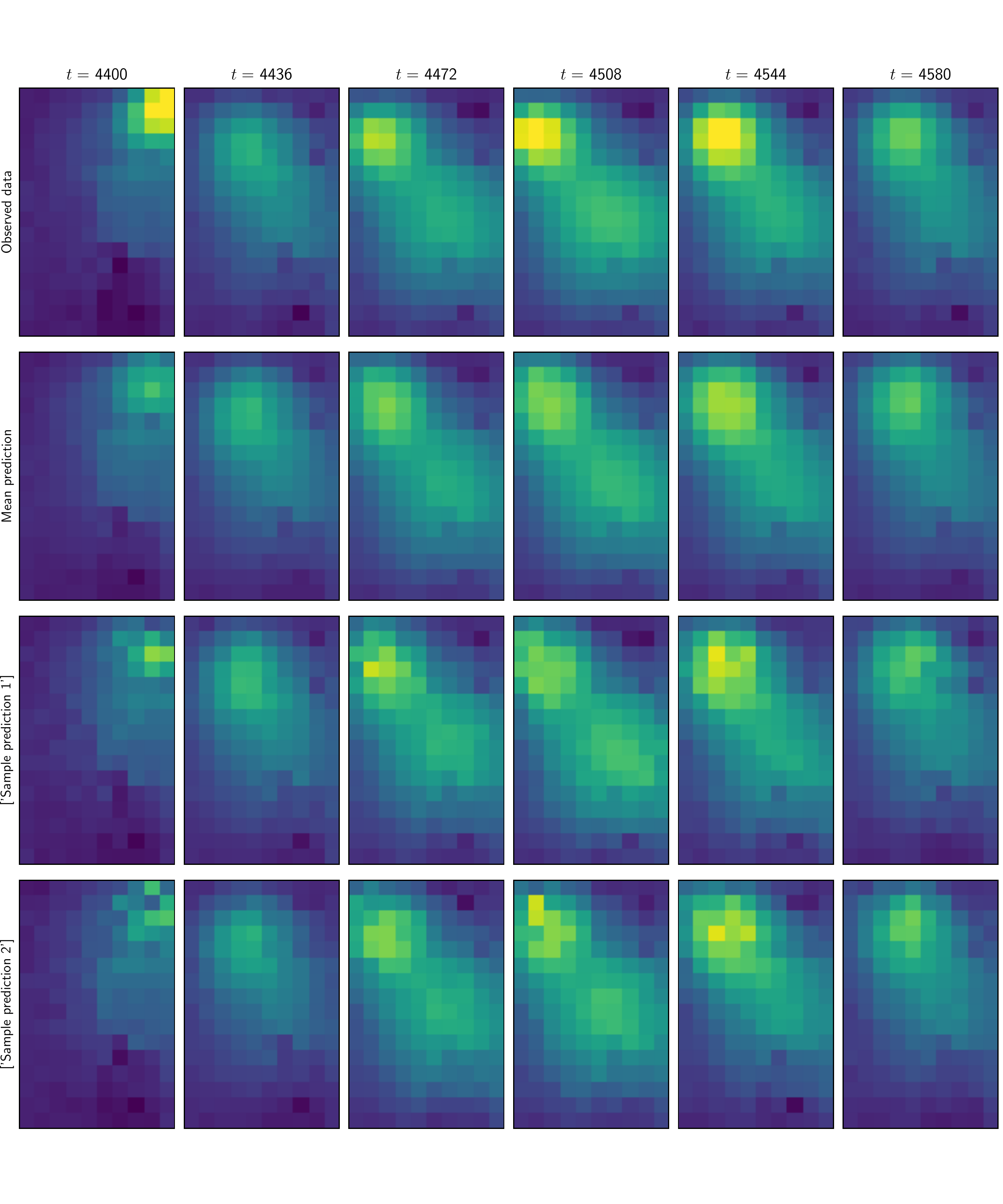}
    \caption{The upper plot shows a time-series of finger poses from the test set, and the lower plot shows samples of the true measured capacitive sensed images, and simulated data from the forward model.}
    \label{fig:TSFWresponse}
\end{figure}

It is interesting to compare forward model performance with and without the availability of simulated data to augment the multi-fidelity model. In Figure~\ref{fig:transferHuman} we show the impact on the CVAE forward model of using only real data, or of using both simulated and real data. For smaller training sets the simulated data has no clear advantage, as the model is still a very poor fit, but as the model starts to reproduce the outputs more reliably, from ca 10,000 data points onwards, the inclusion of simulated data shows a clear benefit. To give a sense of the signals expected in more typical use, Figure~\ref{fig:TSFWresponse} shows a time-series of finger poses, and the associated sensor readings and network predictions.

%Comparisons:
%\begin{itemize}
   % \item We demonstrate the performance of models trained on increasing sizes of training sets, to the impact of the use of the simulation model on the need for comprehensive training data. 
 %   \item Show how little real data you need to calibrate artificial model.
  %  \item Show it adapting to a specific user.

%\end{itemize}

\FloatBarrier
\subsection{Learning the Inverse model}
Once the forward model has been trained, we can use it in the VICI framework to train an inverse model. The forward model can be used to generate extra pairs of sensor and pose mappings to enhance the inverse learning problem. 

We used the same data sets as were used in the forward modelling task, and as there, all results below are on test data which were not used in training.  For the human-generated data with the real sensor, 45205 points were used for training and 31905 points from completely independent runs for testing (care must be taken not to subsample individual runs, due to the high correlation between neighbouring samples).
%For the robot task this **,*** for training and *,*** for test. 

\subsubsection{Network architecture}
The architecture is, similarly to the forward model, a multilayer feedforward model with Leaky PReLU nodes and dense weight layers, and the number of units in each layer is optimised by Bayesian optimisation with ax. 

%\subsubsection{Inverse modelling results on robot-generated data}
%The robot generated task is somewhat more straightforward and repeatable, so gives a good starting point for evaluation. 

\subsubsection{Inverse modelling results on human-generated data}
We compare the VICI approach to inverse model prediction with a direct inverse obtained by training a CVAE network to map sensor images directly to poses, without using a forward model in the process. We used the same architecture for both approaches.

\begin{figure}[htb!]
    \centering
    \includegraphics[width=0.9\linewidth]{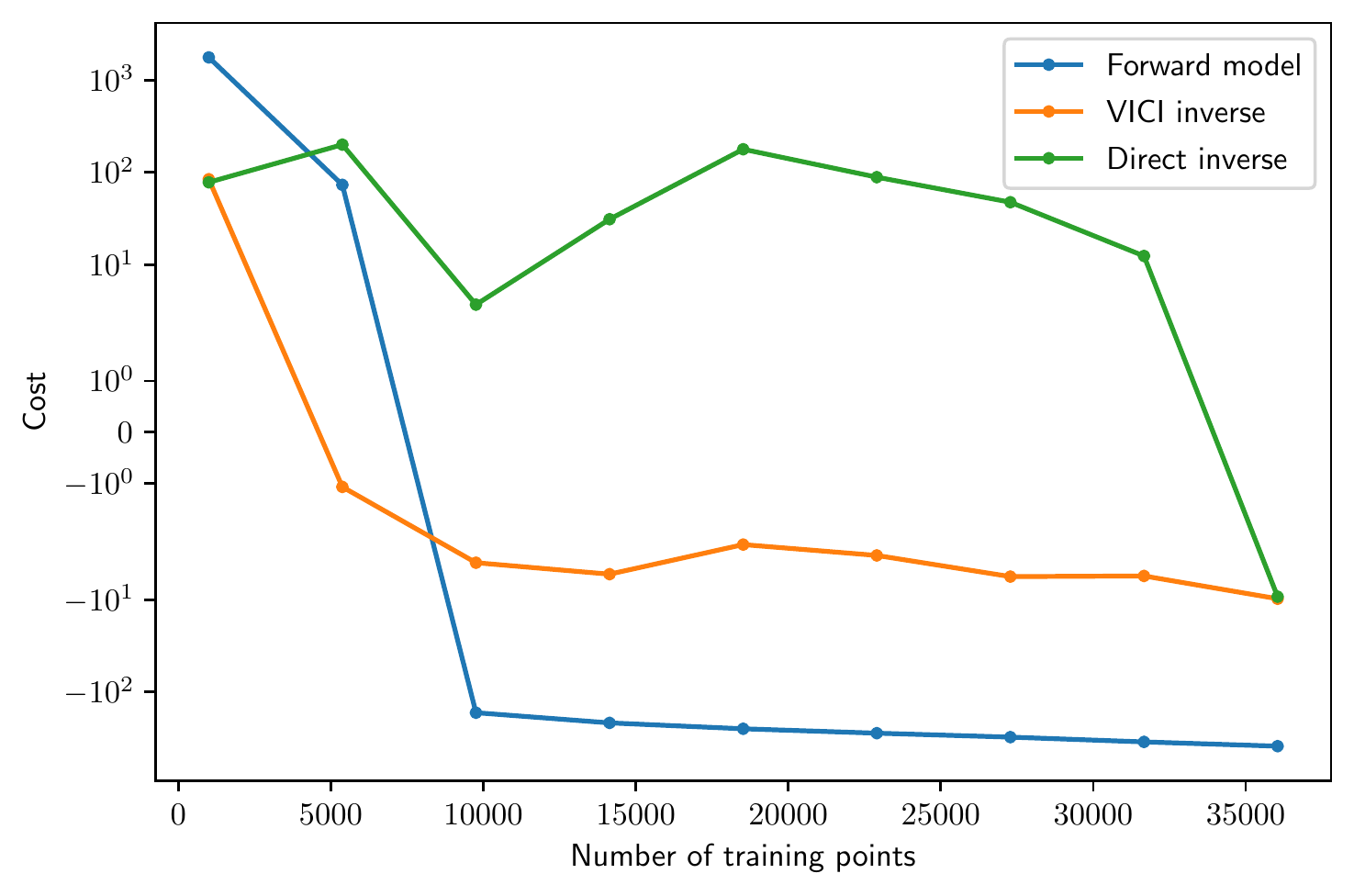}
    \caption{Performance of VICI compared to Direct inverse CVAE for different sizes of training data set. Performance plotted is the negative evidence lower bound (ELBO) cost on the full test set.}
    \label{fig:VICIvsDirectInv}
\end{figure}

Figure~\ref{fig:VICIvsDirectInv} shows the change in performance from very small training sets to the full training set.  Performance plotted is the negative evidence lower bound (ELBO) cost on the full test set (see e.q. (16) in \cite{TonRadTur20}). At each point a forward model was trained using only the limited amount of training data, and the simulation model. The VICI inverse was then trained using the forward model and the limited training data, and a Direct inverse was trained using only the limited training data. As can be seen in the figure, the VICI approach performs better for the smaller training sets, but as the amount of data increases, the advantage reduces, and generally the direct inverse is less robust than the VICI, sometimes performing well, but often with more outliers.

\begin{figure}[htb!]
%\centerline{\includegraphics[width=\linewidth]{xyerrvz.pdf}}
\centerline{\includegraphics[width=0.8\linewidth]{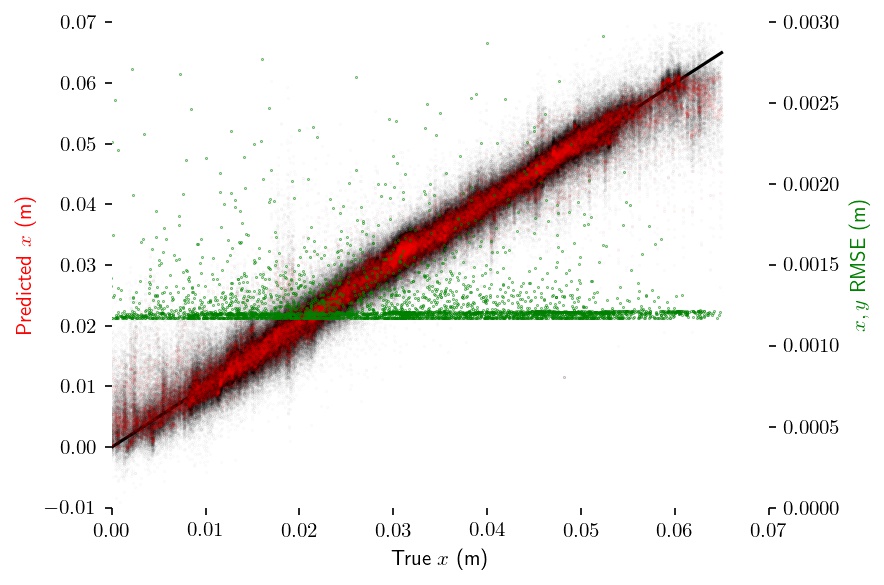}}
%\centerline{\includegraphics[width=\linewidth]{yestvytrue.pdf}}
\centerline{\includegraphics[width=0.8\linewidth]{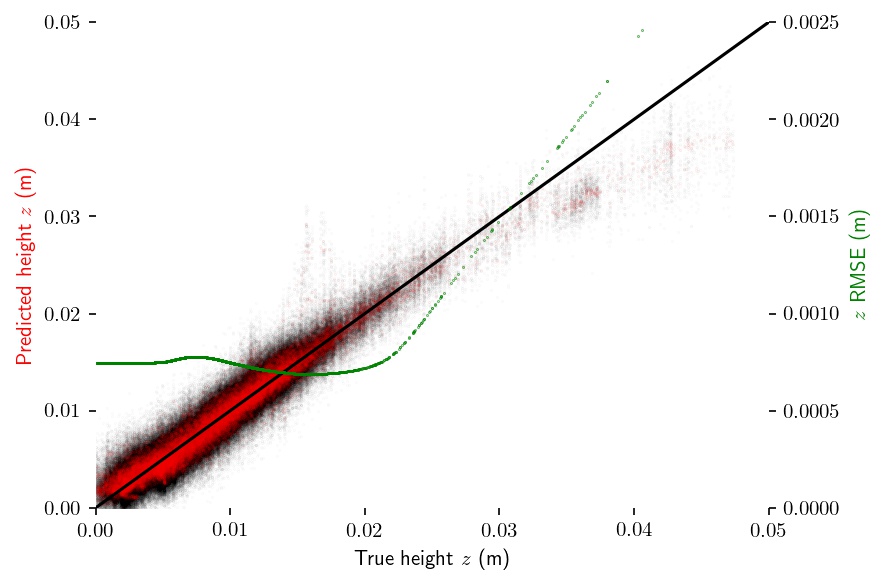}}

\caption{Prediction results on test data for $(x, y, z)$ positions for the Deep network. These show the scatterplots of simulated and real testdata, and the smoothed estimates of the RMSE of each variable over its range. }
\label{fig:estvtrue}
\end{figure}

\begin{figure}[htb!]

\centerline{\includegraphics[width=0.7\linewidth]{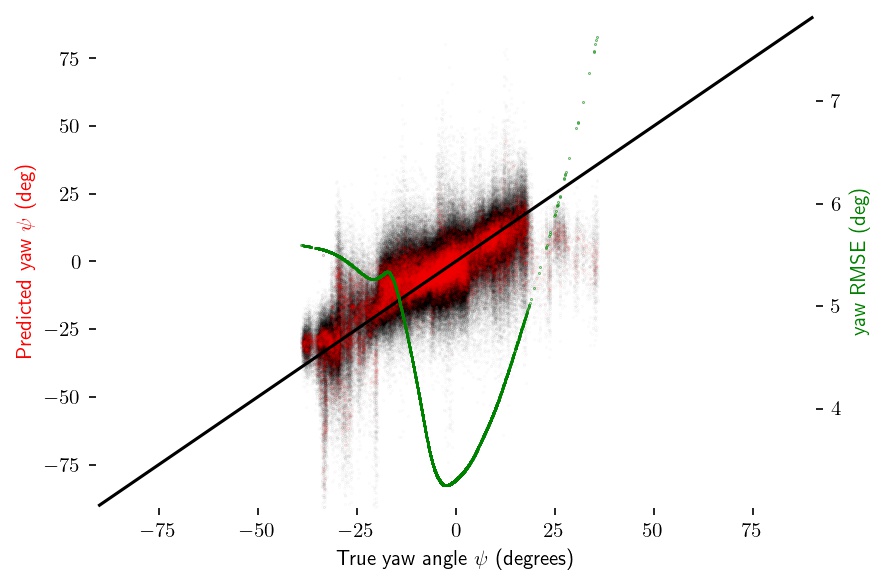}}
\centerline{\includegraphics[width=0.7\linewidth]{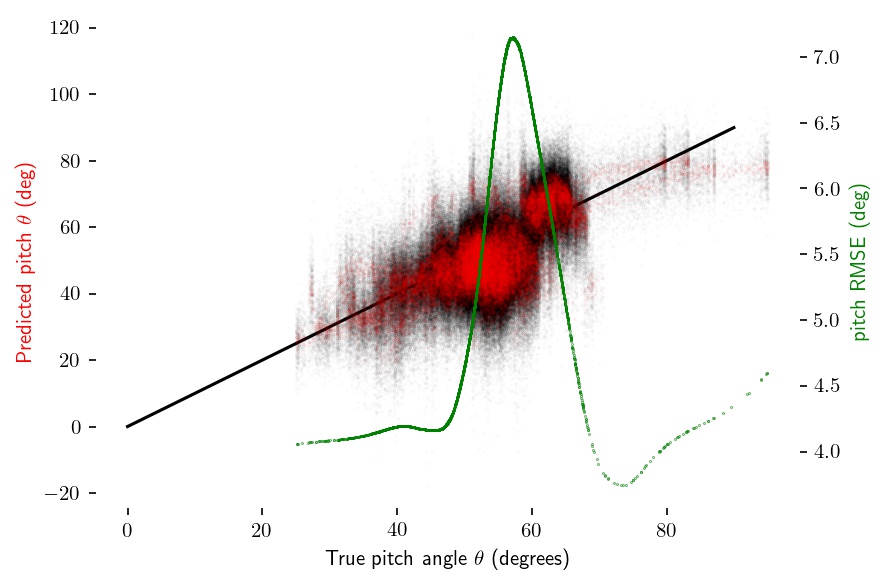}}
\caption{Prediction results for  pitch and yaw angles for the Deep network, combined with RMSE on the test set over their range of angle for both simulated (blue) and real images (green).}
\label{fig:anglevtrue}
\end{figure}

\begin{figure}[htb!]
    \centering
    \centerline{\includegraphics[width=0.69\linewidth]{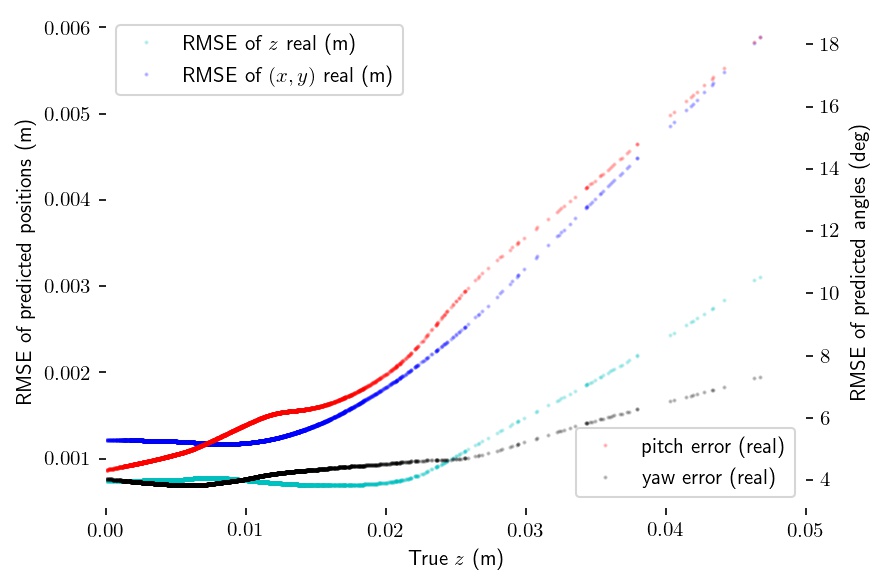}}
    \caption{Smoothed estimates of the RMSE of each variable as $z$ varies, for both simulated and real results.}
    \label{fig:errvz}
\end{figure}

We now present more detailed results on the VICI approach. Figure~\ref{fig:estvtrue} shows the scatterplot of predicted vs measured $(x, y)$ and $z$ positions. These show good general correlations, with height predicted well up to 5cm from the device. The outliers tend to to be associated with larger heights, as shown in Figure~\ref{fig:errvz}.  Pitch and yaw have traditionally been more difficult to infer for capacitive screens, especially as $z$ increases. In figure~\ref{fig:anglevtrue} we see very strong correlation between predicted and measured pose angles, with the VICI approach. The variation of the errors with $z$ for all of the inferred positions and angles is summarised in Table~\ref{tab:zrangeerr}. Figure~\ref{fig:errvz} provides a visual representation of how the error increases as $z$ increases. The overall $x,y$ RMSE was 0.21cm, the RMSE on $z$ was \SI{0.13}{\cm}.  
The overall pitch RMSE was \ang{7.8} and yaw RMSE was \ang{7.2}. 

\begin{table}[htb!]
    \tbl{Errors at different heights with VICI}{
    %\caption{RMSE of the different inferred pose elements at different heights}
    \centering
    \begin{tabular}{c|l|l|l| l}
    {\bf Distance range (cm) \#} & {\bf $x,y$ RMSE} & {\bf $z$ RMSE}  & {\bf Pitch RMSE (deg)} & {\bf Yaw RMSE (deg)}\\
0 - 1 & 0.18 & 0.11 & 6.9 & 6.5 \\
1 - 3 & 0.19 & 0.12 & 7.7 & 7.1 \\
3 - 5 & 0.21 & 0.12 & 7.8 & 7.0 \\
5 - 10 & 0.21 & 0.13 & 7.9 & 7.1 \\
all& 0.21 & 0.13 & 7.8 & 7.2 \\
    \end{tabular}
    }
    \label{tab:zrangeerr}
\end{table}

%Show how performance varies depending on the amount of data available to calibrate the artificial model.

%Show the impact of including the simulation model.

%\begin{figure}[htb!]
%\centerline{\includegraphics[width=1.1\linewidth]{pyerrvz.pdf}}
%\centerline{\includegraphics[width=1.1\linewidth]{xyerrvz.pdf}}
%\caption{3D plots shows the correlations of $\theta, \psi$ (left) and $(x, y)$ (right) errors with increases in $z$}
%\label{fig:xyerrvz}
%\end{figure}


%\subsection{Analysis of network features}
%What has the system learned? We can perform saliency analysis to find the optimal inputs for each of the pose states.

%\subsection{Impact of covariance among outputs on learning}
%The pitch and yaw of a finger will affect the inference of $(x, y, z)$. To demonstrate this, we tried to learn a model which was only trained on  $(x, y, z)$, with no pose data. However, this did not change the accuracy of the predictions on $(x, y, z)$, suggesting that the representations used to infer pitch and yaw are not making sufficient difference to improve location estimates.... REMOVE?

%\subsubsection*{Examples of uncertainty prediction}

%Does the model uncertainty predict this?
%Look at high pitch angles and yaw
%Look at uncertainty as z increases
%Look at performance of the inversion in areas with less dense training data, with and without simulated data.

\FloatBarrier
\section{Generalising to new users}
So far we have taken data from a single user to explore the modelling capability of this approach, combining a physics-based electrostatic model with observations from human use, to give us an approximation of the forward model. We can now apply the same principle as before, and use our combined forward model as a starting point for calibration to a new user's data. We would like to answer the questions: How well does the model perform out of the box on a new user, without having trained on any of that user's data? How much data is needed to re-calibrate a previous forward model to a new user?

\begin{figure}[htb!]
    \centering
    \includegraphics[width=0.49\linewidth]{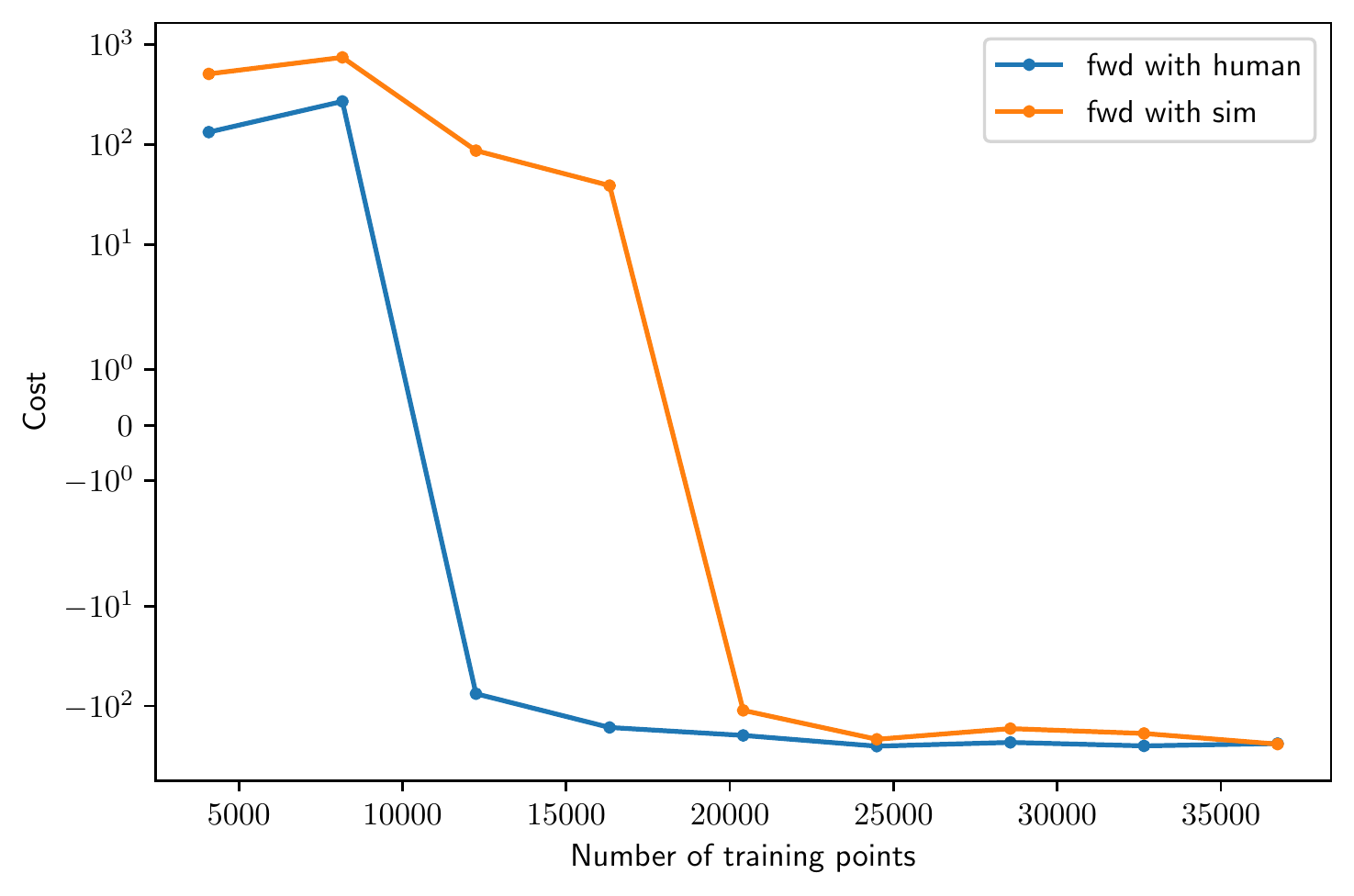}  \includegraphics[width=0.49\linewidth]{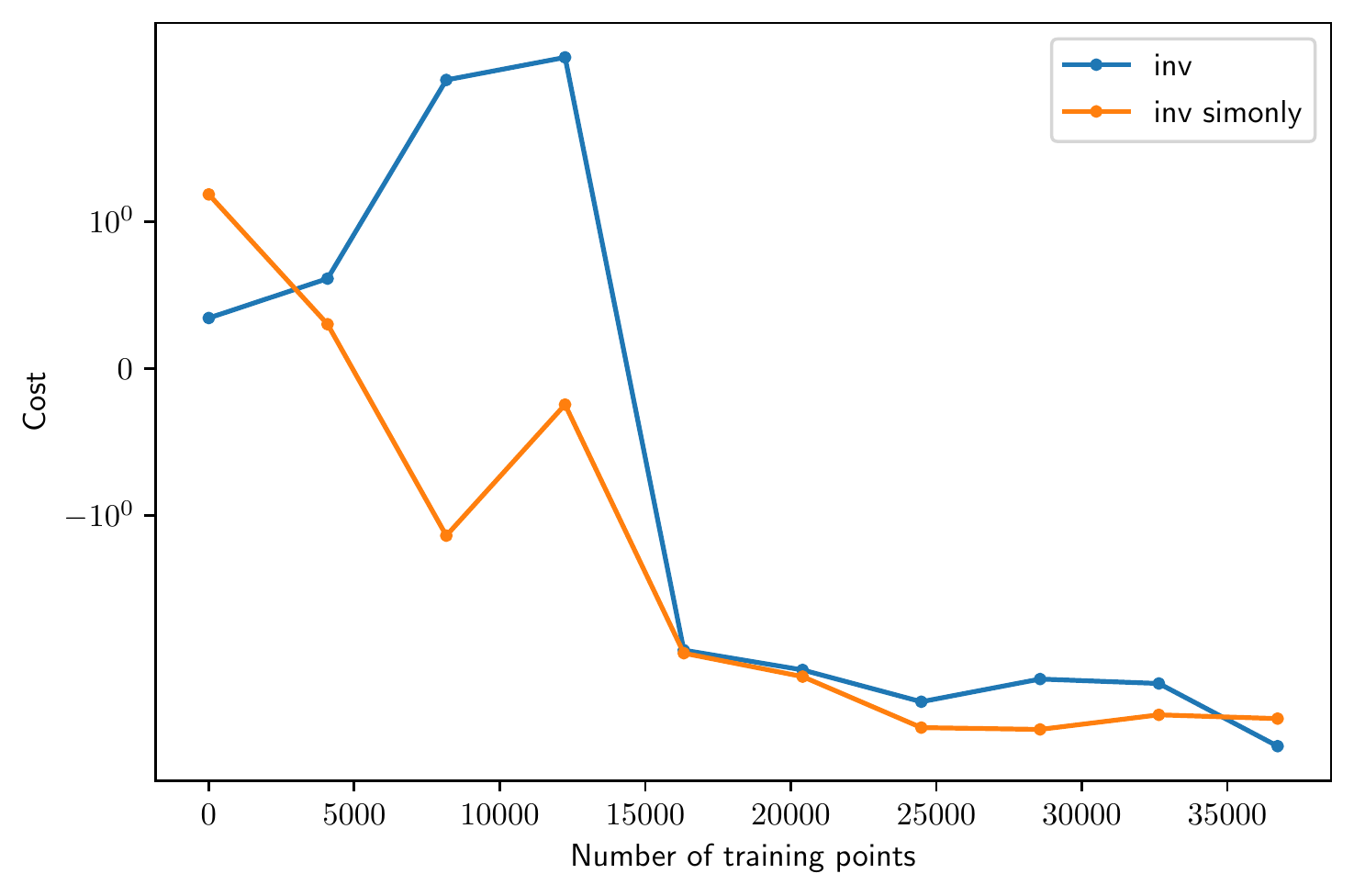}
    \caption{Left: Performance of the forward model for different training set sizes. Comparison of a low-fidelity model with a human-calibrated forward model from a different user. Right:Performance of the inverse model for different training set sizes. Comparison of two multi-fidelity models, where in one case the low-fidelity model is the electrostatic emulator used earlier while the other is a combination of the electrostatic emulator and a human-calibrated forward model fully trained on a different human user.}
    \label{fig:fwdTransferWithoutHuman}
\end{figure}

Figure~\ref{fig:fwdTransferWithoutHuman}, left, shows how the forward model is able to re-calibrate based on different amounts of calibration data, and that the resulting model is more accurate than one which starts from scratch with only the low-quality electrostatic simulation model. Both models are, however, poor matches for small amouts of calibration data.  Although the forward model is consistently more accurate when using the previous user's model as a starting point, Figure~\ref{fig:fwdTransferWithoutHuman}, right, shows the impact of the extra calibration data on the quality of the inverse solution for the case of building on the first person, or just on an electrostatic simulator, and there the behaviour is more mixed, despite the consistently higher performance of the forward model, the simple simulator seems to lead to more robust inverse models until ca 15,000 training examples (although up to this amount of data, all are poor quality models), at which point, the forward model quality improves, and the inverse models improve significantly, but are fairly similar in performance. This suggests that while the invididual model presented in the previous sections shows a high level of accuracy, more careful consideration of multiuser models and training collections is needed, if more flexible, general purpose models are desired.

\section{Time-series analysis}
\begin{figure}[htb]
    \centering
    \includegraphics[width=0.46\linewidth]{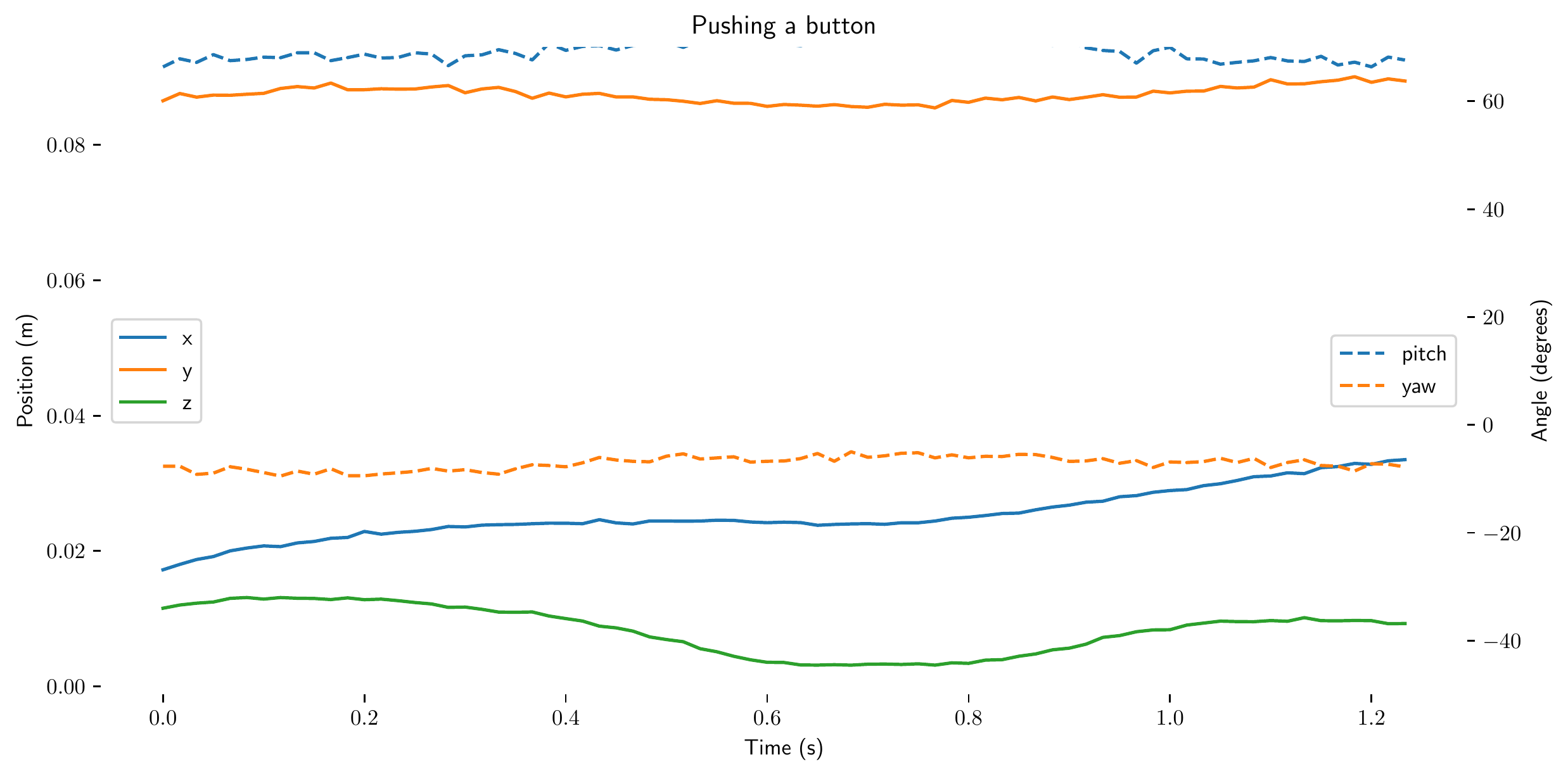}\hspace{2mm}
    \includegraphics[width=0.46\linewidth]{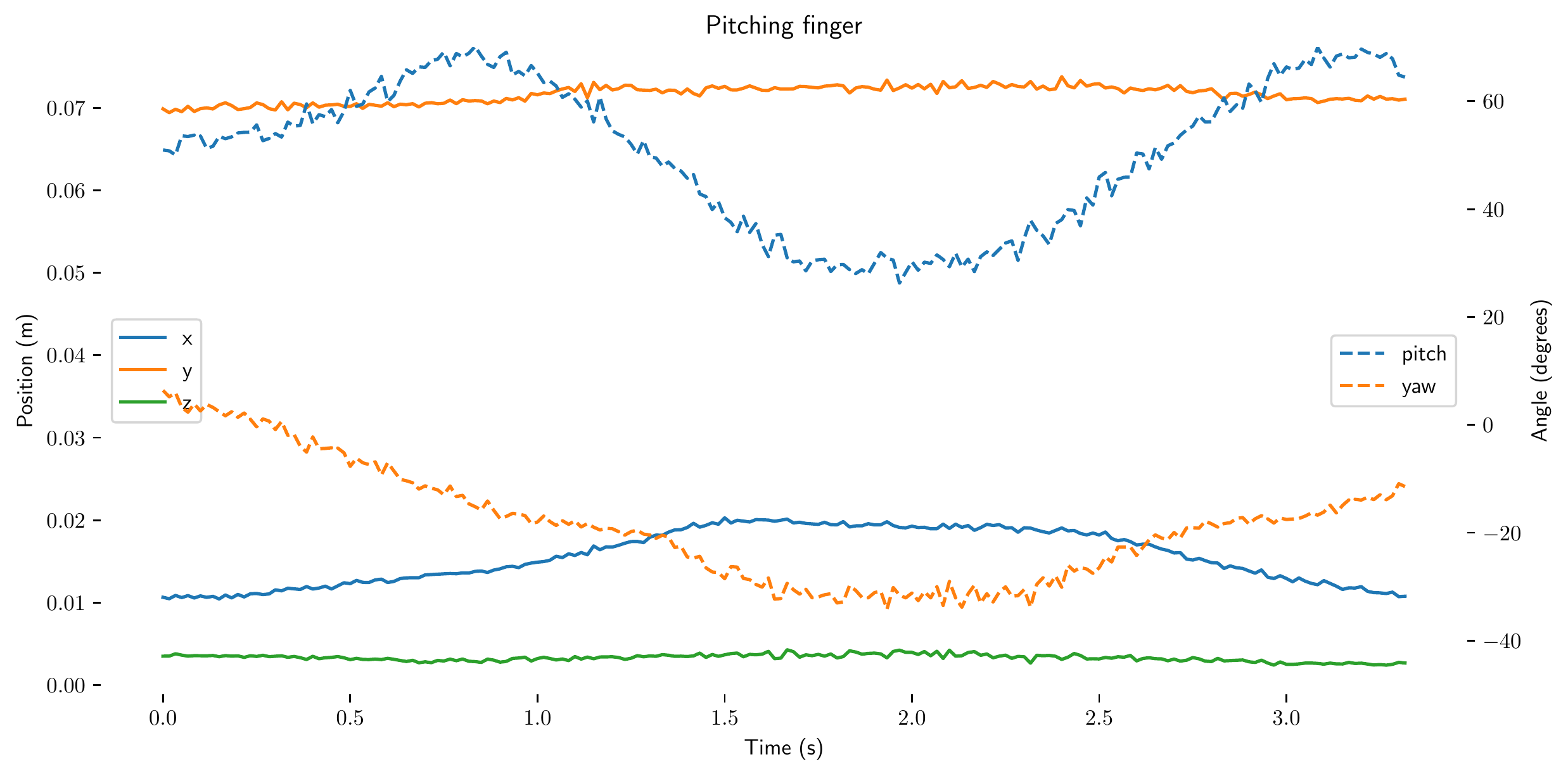}
    \centerline{\hspace{2mm}\includegraphics[width=0.46\linewidth]{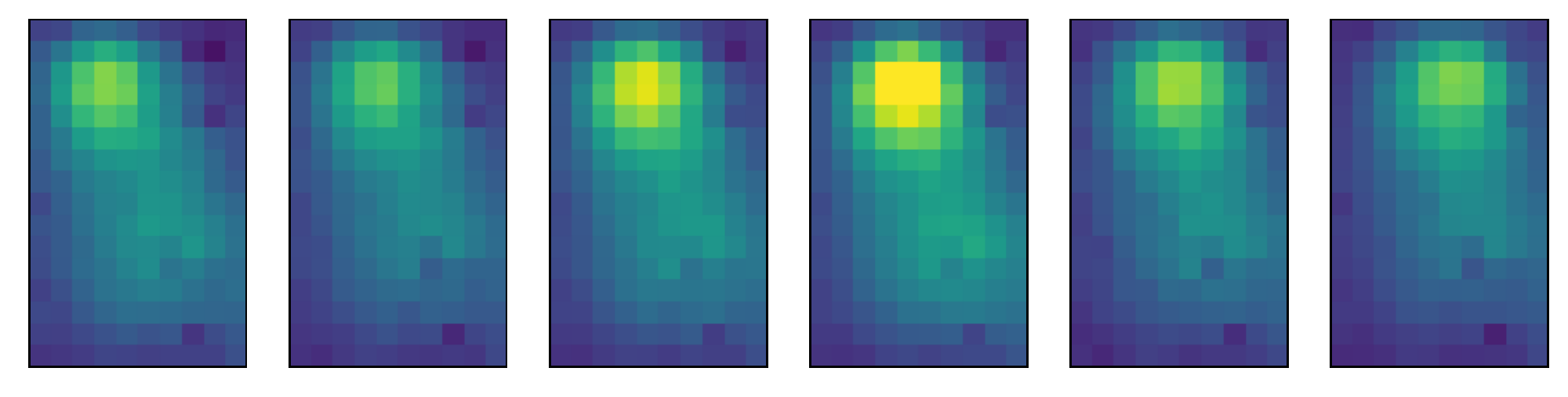}\hspace{2mm}
    \includegraphics[width=0.46\linewidth]{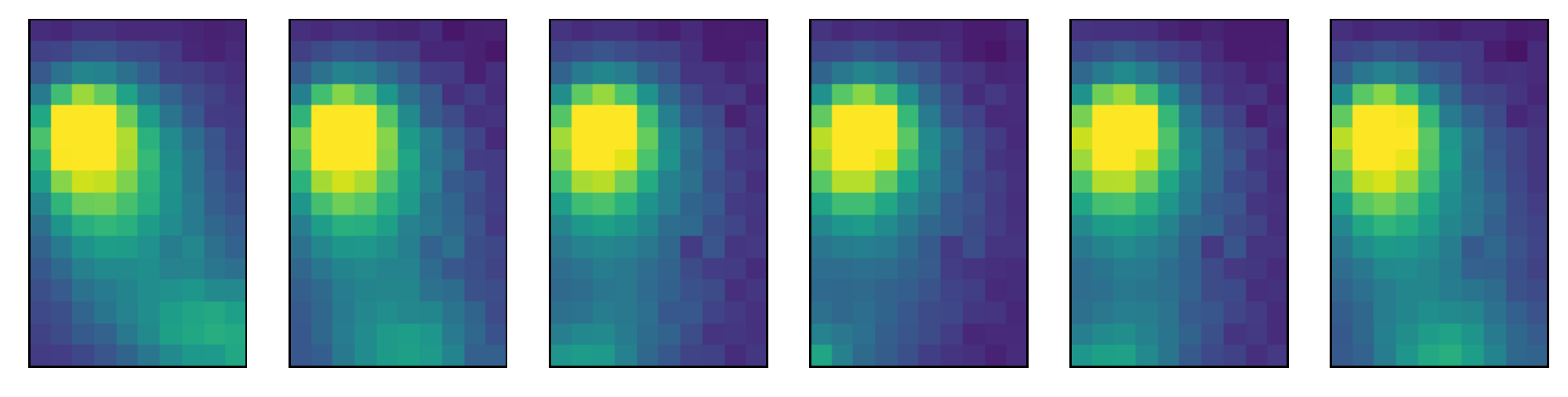}}
    \includegraphics[width=0.49\linewidth]{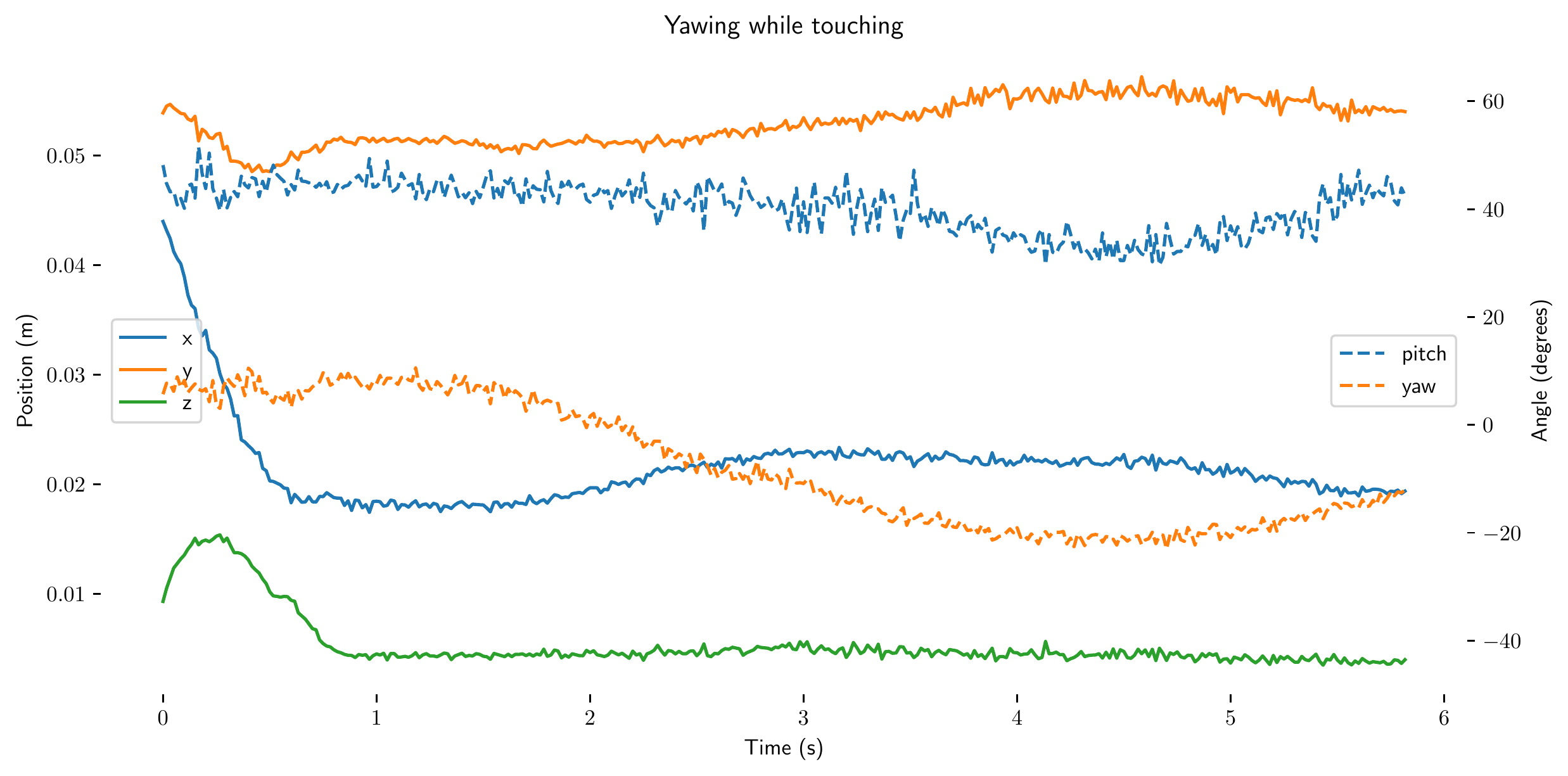}\hspace{2mm}
    \centerline{\hspace{2mm}\includegraphics[width=0.49\linewidth]{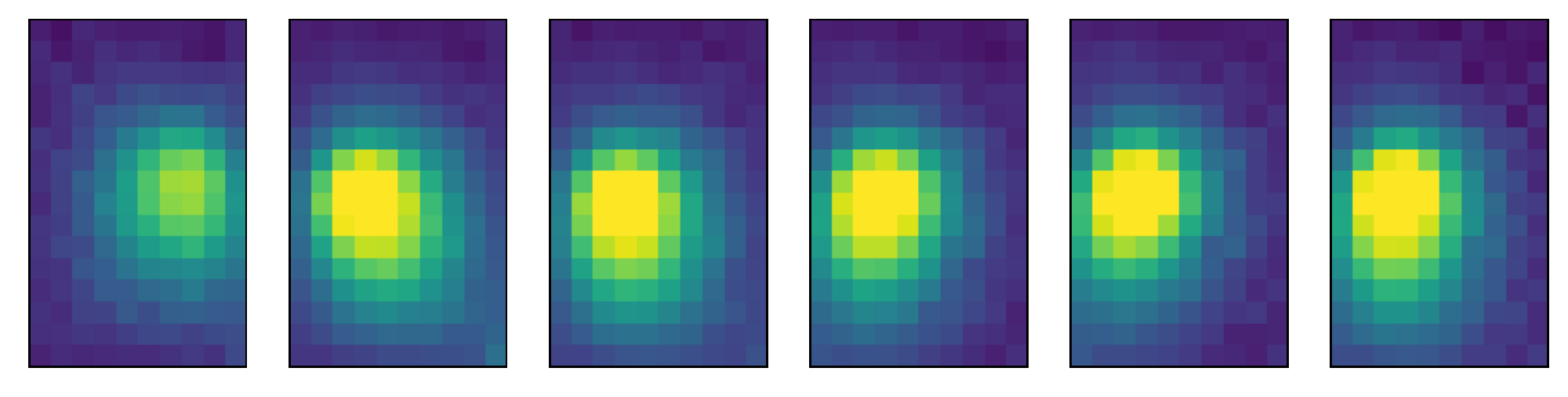}}
    \caption{Several examples of interactions where the pose time-series shown are inferred from the capacitive sensor data shown in the lower part of each figure. Top Left: Pushing an on-screen button from *cm starting position. Top Right: Pitching motion with the finger. Bottom: Yawing motion at *cm above the screen}
    \label{fig:push}
\end{figure}

The previous sections analysed the performance in a static manner. Of course, most touch interaction is dynamic, and the potential benefits of a 3D capacitive sensor relate to movement around and towards the device. We therefore show some examples of time-series plots of movement types below:

In Figure~\ref{fig:push} we show examples of inferred pose time-series from sensor readings for button pushing, pitching and yawing motions. In these examples each prediction is independent of the others, but it would be straightforward to add a filter such as a Kalman filter or particle filter to smooth the results from neighbouring time-points.

\begin{figure}[htb!]
    \centering
    \includegraphics[width=0.49\linewidth]{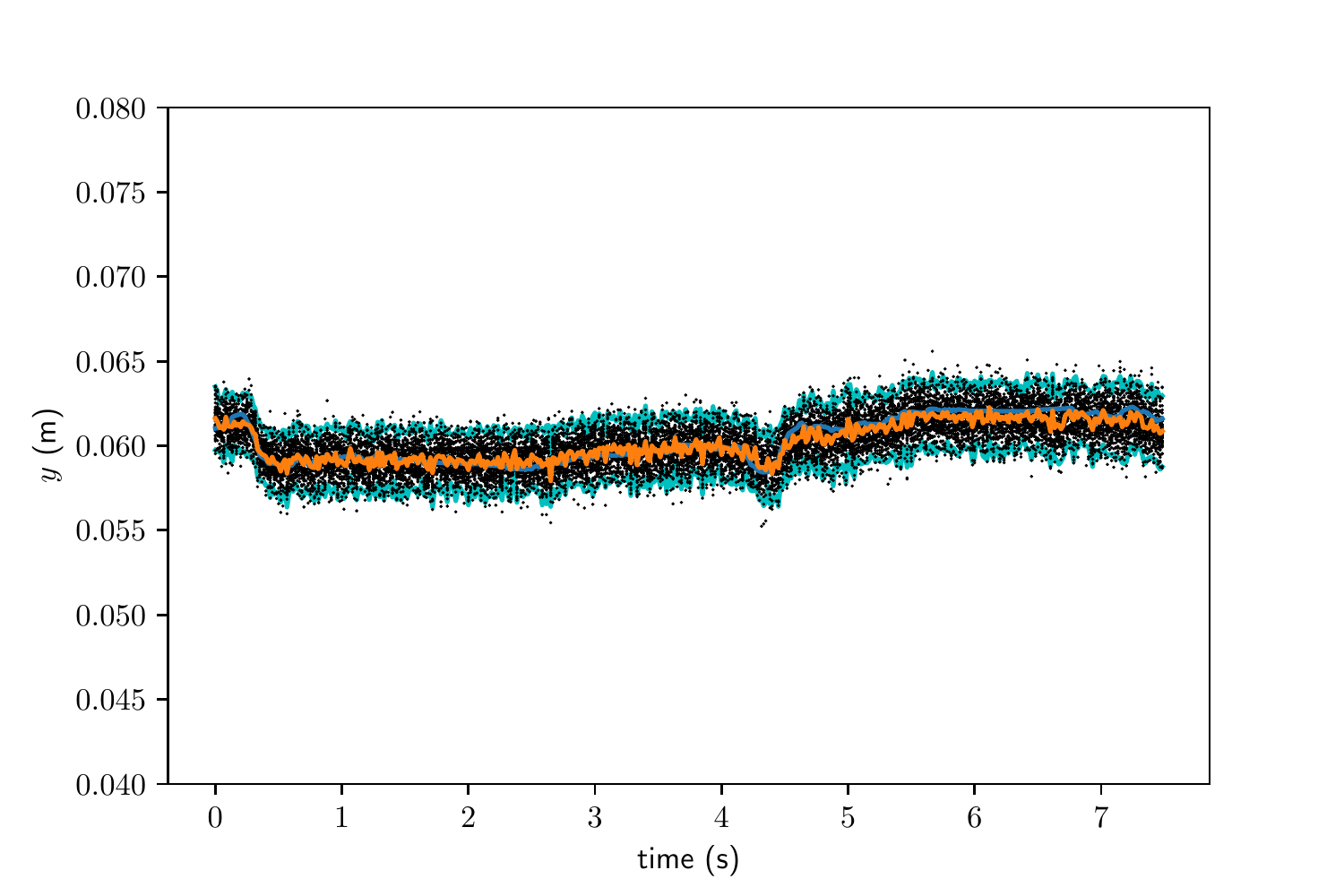}
    \includegraphics[width=0.49\linewidth]{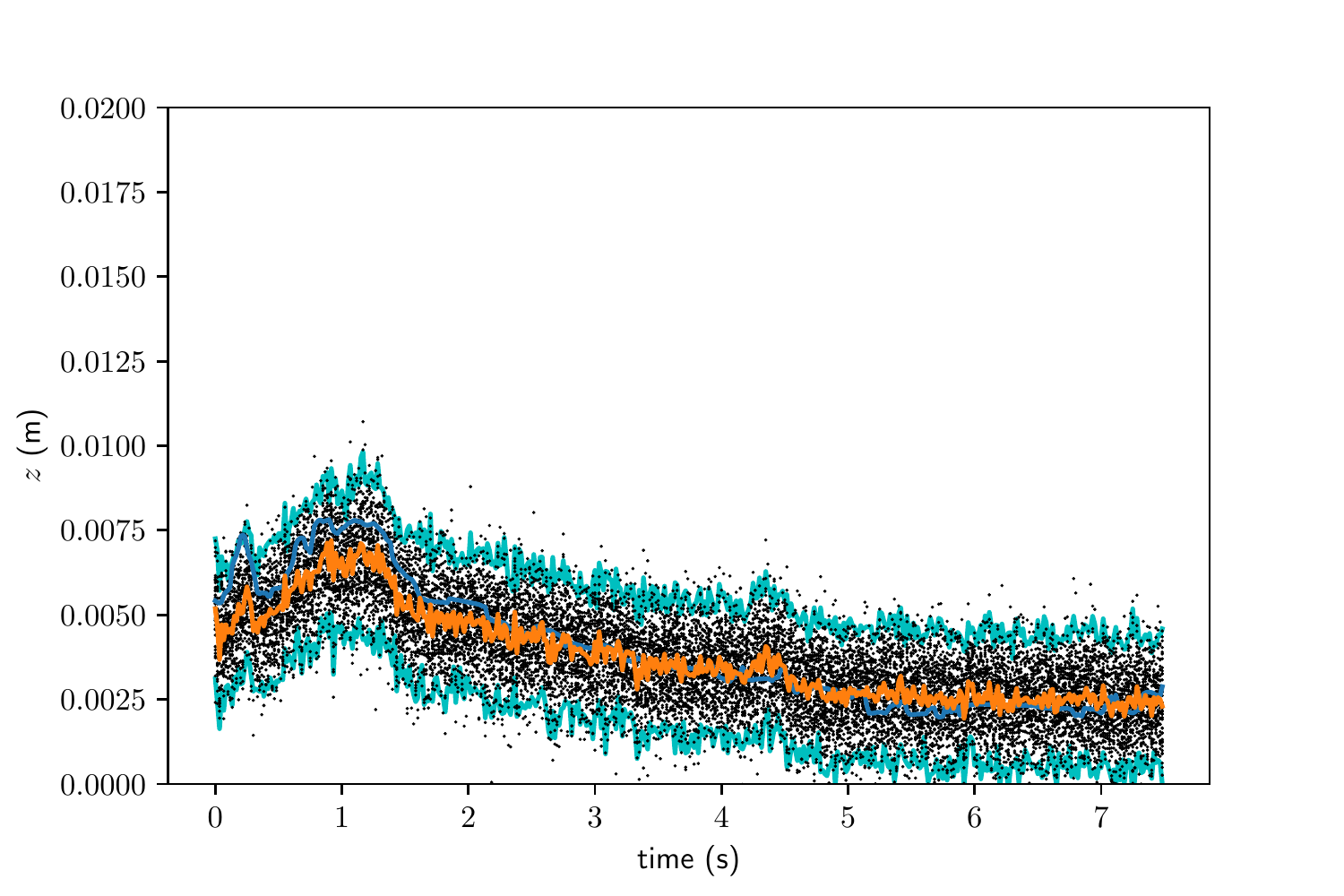}
    \includegraphics[width=0.49\linewidth]{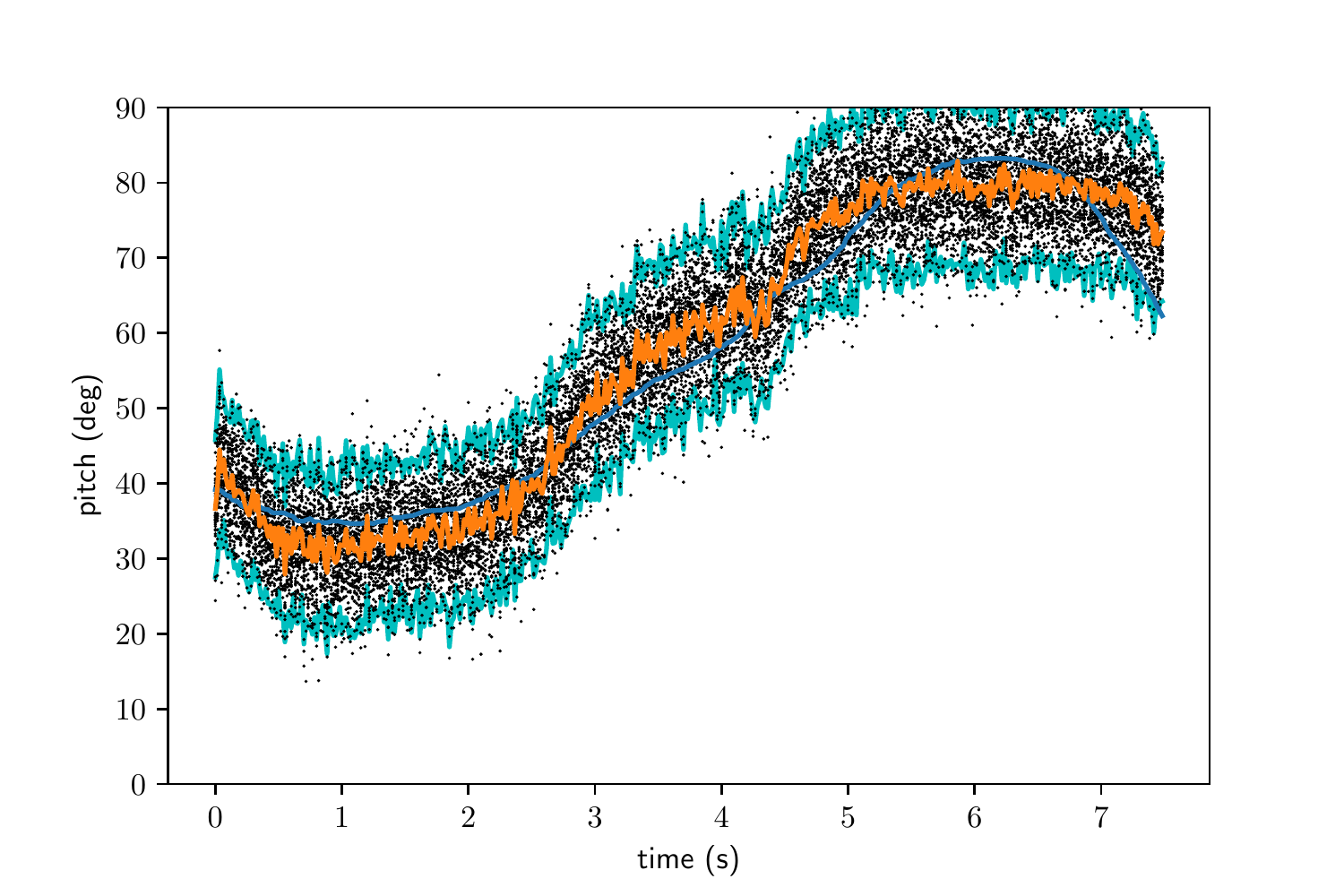}
    \includegraphics[width=0.49\linewidth]{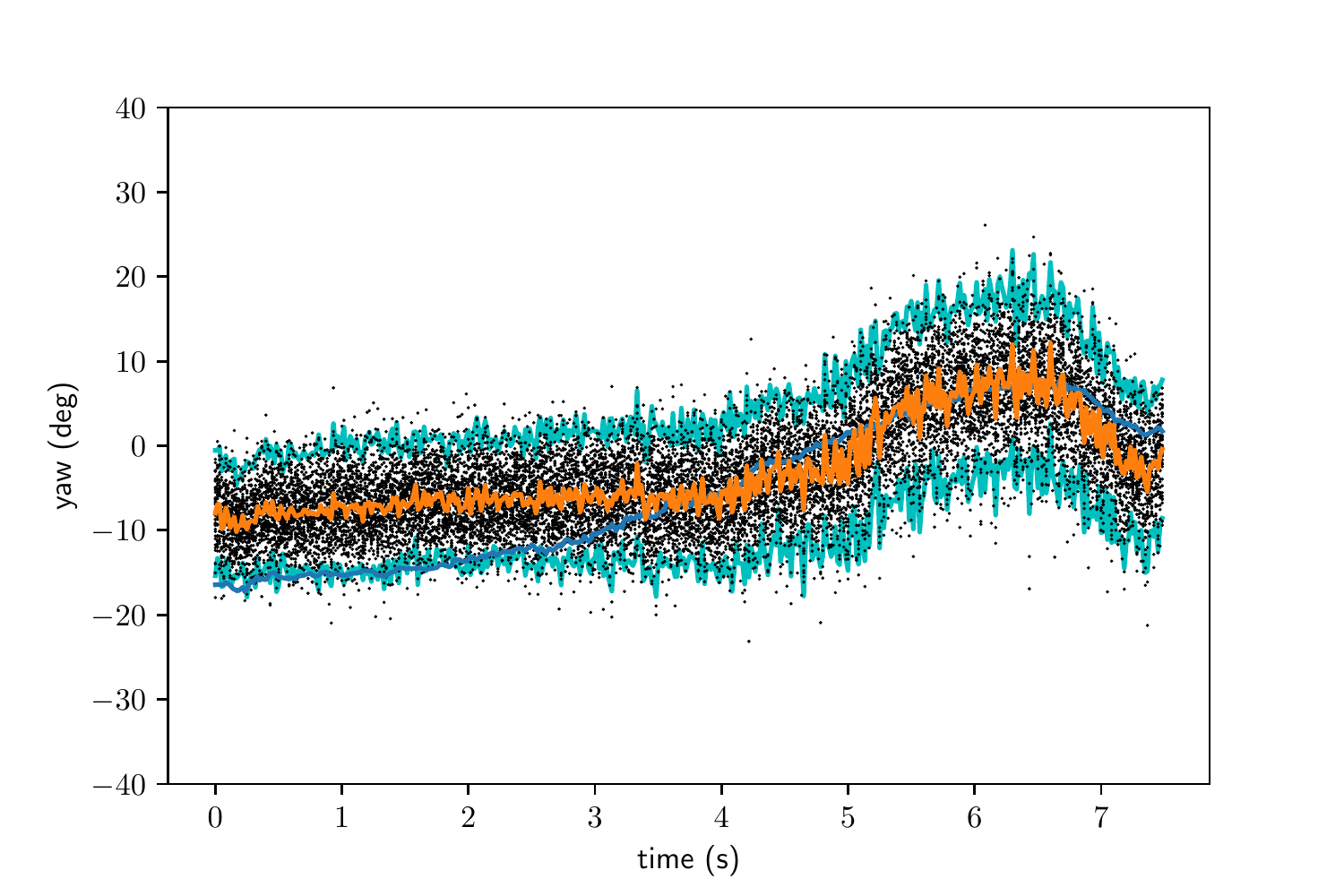}
    \caption{An example of part of a time-series of predictions of pose, from measured sensor data. Upper Left: $y$, Upper Right $z$. Black points are samples, cyan lines show $\pm 2\sigma$, blue lines are the measured states. Lower left pitch, Lower Right yaw. Note how the uncertainty varies, but generally does a good job of predicting the uncertainty.}
    \label{fig:TSuncertainty}
\end{figure}

Figure~\ref{fig:TSuncertainty} shows the variability of the time-series inferences as a movement develops over time. Each of states shows a different level of uncertainty, reflecting how reliably the inverse model can solve it from the sensor data. These are all independent samples, with no attempt to introduce correlations over time, which could smooth out the inferred results significantly. 
\FloatBarrier
\section{Forward \& backward combinations}
The forward (`causal') and inverse (or `regression') modelling approaches can be combined via the use of the one-step regression model to create an initial condition for the optimisation or particle filter approaches to refine and smooth. However, in this paper we leave the prediction process as a single-step one. 

\subsection{Differentiable models}
An advantage of the conversion of general simulation code to a deep network running on a Tensorflow graph is the ease with which we can analytically calculate the model gradients -- a simple call to \lstinline{tf.gradients(loss, model.input)} suffices. This is useful, e.g.  when inferring which inputs might have caused an observed sensor vector. Starting from a random initial guess, $x_0$, and adapting  $x \propto \frac{\partial L}{\partial x}$, for our loss function $L$, we can converge on a plausible input vector $x$. %Figure~\ref{fig:inp_opt} shows such an optimisation process.

%\begin{figure}[htb!]
%    \centering
 %   \includegraphics{}
 %   \caption{Before and after input optimisation}
%    \label{fig:inp_opt}
%\end{figure}

\subsubsection{Detecting anomalies}
An advantage of probabilistic models, such as the CVAE models used in this work is that they can return multiple samples, allowing us to calculate statistics such as the covariance as well as the mean of their predictions, allowing the system to infer when the results can be trusted. 

A further validation can come from a two-stage approach from observed sensor readings via pose estimate to predicted sensor readings, we allow the system to automatically generate a test of  prediction accuracy from the regression-based model. If the inferred pose and position vector generates a poor match to the observed sensor reading, we might be outside the area trained (possibly due to disturbances, such as other fingers), and the results are uncertain. This can also be used to automatically generate new experiment points to augment the training data (or suggest poses for the user to generate). In cases where the forward model prediction of the real and the simulated results deviate, we have evidence for a `non-optimal' input (e.g. in the touch case, other non-pointing fingers may be dragging below the hand). Our approach allows us to both infer the most likely position, but also provide feedback that the user could improve precision by changing their hand pose.

\FloatBarrier
\section{Simulation-based training}
Using the electrostatic simulator allowed us to generate much larger datasets, compared with the use of the specific robot system used by our industrial partners, which required frequent manual intervention and recalibration. This was due to a substantial decrease in the time required to simulate a single point (ca. 80 seconds on an Intel Core i7-6700K 4.0GHz Quad-Core Processor, which can be  parallelised on multicore machines), and the ability to run large batch jobs of points as a background process with no need for manual intervention. This has potential advantages:
\begin{enumerate}
\item It makes it feasible to generate datasets of thousands of points at a controlled range of positions and poses, in a relatively short period with minimal effort from the experiment designer, and in a completely controlled manner, compared to using human test subjects.
\item While the measurements of the capacitive sensors are not identical, the approach eliminates measurement errors on the position and pose of the finger, whether from a tracking system or robot sensors.
\item Simulation studies enables testing of system configurations and sensor algorithm design even in the absence of finished sensor hardware, providing the basis for a process for device manufacturers to follow to predict how different design decisions could affect the performance of the system and to create reliable 3D touch screen interaction.
\end{enumerate}
Our immediate goal was to explore whether machine learning could improve our ability to predict yaw and pitch in non-contact areas between \SIrange{1}{3}{\cm} from the screen, and the results from the simulator gave us a qualitatively similar level of complexity, and acted as a `best case' scenario. The use of simulation let us rapidly generate particular combinations of interest, and test, compare and adapt machine learning model structures to the task in hand without the complexities of human trials and the associated sensing uncertainties.

\subsection{Transfer learning:} In many machine learning tasks it is possible to get a lot of data for a closely related task. The forward simulation approach  has the advantage that we can simulate the capacitance for an ideal sensor without disturbances due to sensor noise. This is potentially useful as part of the representation learning process, as it can help the forward model generate the appropriate structures. Furthermore,  we can simulate a wide range of finger types, poses and positions, but the simulated model will always be an approximation of the real sensors. We experimented with different proportions of human-generated and simulated data. In terms of the loss function for human-generated test data, the forward model benefitted from simulation data when small sets of human-generated data were available, but as human data increased further the performance gap reduced.

\subsection{Accelerated and differentiable simulation}
\begin{figure}
    \centering
    \includegraphics[width=\linewidth]{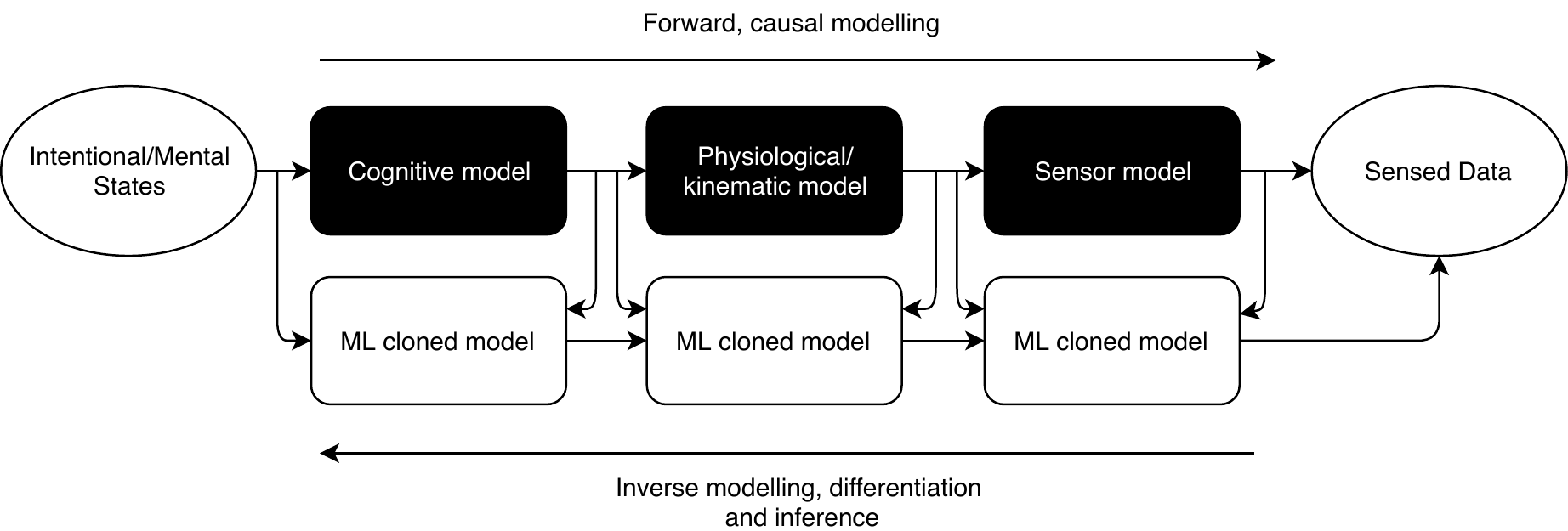}
    \caption{Concatenated series of differentiable models, using them to solve forward and inverse problems in HCI contexts.}
    \label{fig:concatmodels}
\end{figure}

The use of common ML frameworks permits a computationally efficient, accelerated implementation of the forward, or causal model implemented by the electrostatic simulation (or from physical experiments)  which is also algorithmically differentiable. This can then be used in real-time to infer the most likely inputs, by performing gradient-based optimisation to minimise the distance between the current sensed values and those of the model output. It also means that, as shown in Figure~\ref{fig:concatmodels}, we can concatenate a series of simulation tools, `clone' them with machine learning tools and then efficiently differentiate the models for optimisation or control purposes. This  simulation pipeline for the sensor hardware can be augmented with biophysical models of human movement, e.g. \cite{BacPalOul15} and \cite{Oul18a}. In real use, the dynamics of the finger movement will be constrained by such biomechanical aspects of the human hand, and highly correlated with errors. In conventional touch screens this would only be of use for two-dimensional dynamic gestures, but with the depth sensing in this system, we can also acquire and analyse multiple frames of the image as a finger approaches the screen. This provides extra information which can help disambiguate the pose and position, and improve tracking over time. 

\section{Conclusions}

\subsection{Inverse models in HCI methodology}

 Recurring questions in Computational HCI design include: {\it `How can we go from some sensor readings to knowledge of the human physical state?'} and {\it `How can we go from the physical state to the associated user intention?'}. These questions will then interact with questions such as: {\it How does the variability among different users, or different performances of the actions map into variations in physical performance, and hence sensing, and via inversion, back to the inferred intention?} In other words, the inversion process may be robust to some variations in the forward process, but very sensitive to others. If the inverse problem for a particular task and interface combination is difficult to solve, leading to a broad posterior distribution, it means that the interface is likely to become difficult to use, because the system cannot reliably infer an intention from the sensed data. 
 
  Systematic application of probabilistic forward and inverse models has the potential to turn the analysis and design of input systems for human--computer interaction from an art into a science, putting the inference of human intent from sensed signals onto a much more rigorous basis.  
 
 We suggest that a key goal of the design process is to computationally design interfaces to make them easy to invert; designing the sensor characteristics, signal processing and interaction metaphors, such that smooth, natural, predictable interactions on the part of a user correspond to efficient tours of the information space, augmented with rich, human-centric feedback. We can design systems such that tools that humans are good at controlling can generate intention signals that are easy to reliably identify, and we can do so in simulation.  
 
In this paper, we demonstrated on a real-world, challenging task how theoretical models and physical observations could be combined with modern machine learning models to create leading-edge performance, based on accurate differentiable and invertible models.

\subsection{Simulation approach to system design}
When designing novel sensor hardware it would be advantageous to be able to create a simulation pipeline from which we can predict the performance of different physical sensor systems and the associated algorithms for inference of intention from human movements.

The use of the electrostatic simulator allowed us to flexibly explore a range of use scenarios, and to experiment with assumptions and design trade-offs. The simulator was made computationally feasible for real-time use by emulating it from data with a neural network. We were then able to use it to enhance the multi-fidelity forward model, which could improve robustness of the inverse mapping, inference pose from sensor information. 

\subsection{Demonstration of 3D pose inference for capacitive sensing}
The machine learning methods used in this paper, along with the novel hardware and the use of multiple approaches to generating training data, including simulated and human-generated sources, allowed us to improve accuracy of inference of 3D finger position and finger pose for capacitive screens on a mobile device. We demonstrating accurate position and pose inference at a range of heights and positions around the device.

The RMS error on $z$ from 1.2mm in the 1-3cm range, and 1.9mm on $(x,y)$. Errors in $(x,y)$ were from 1.8mm in the 0-1cm range, and increased with increasing $z$ to 2.1mm above $z$=5cm. Pitch and yaw had higher RMSE rates of \ang{7.7} for pitch and \ang{7.1} for yaw in the $z \in 1-3$cm heights, but interestingly, did not get much worse with height, \ang{7.9} and  \ang{7.1} for $z>5$cm.  For comparison, the best pitch and yaw results to date for cases where the finger is touching the surface were reported in \cite{MayLeHen17}, for commercially available screens, as  RMSE of \ang{12.75} for pitch and \ang{24.48} for yaw, whereas our results for RMSE on contact are \ang{6.9} for pitch and \ang{6.5} for yaw.

%The sensor cannot determine variation in yaw angles when the finger is perpendicular to the screen.

These suggest that although pitch and yaw information can potentially be used even when the finger is not in contact with the screen, that the accuracy is lower, so  more coarse partitions in pose space would be used. The probabilistic ML methods used enabled us to estimate the uncertainty of the solution of the inverse problem, which can be of further novel use in interaction design, and can help in appropriate design of variable coarseness quantisation of poses. It is also interesting to note that we were able to make good predictions beyond the edges of the screen, suggesting the possibility of interaction modes above the sides, where the finger would not obscure the screen. 

\subsubsection{Research implications of 3D touch}
\begin{figure}[htb]
\centerline{\includegraphics[width=\linewidth]{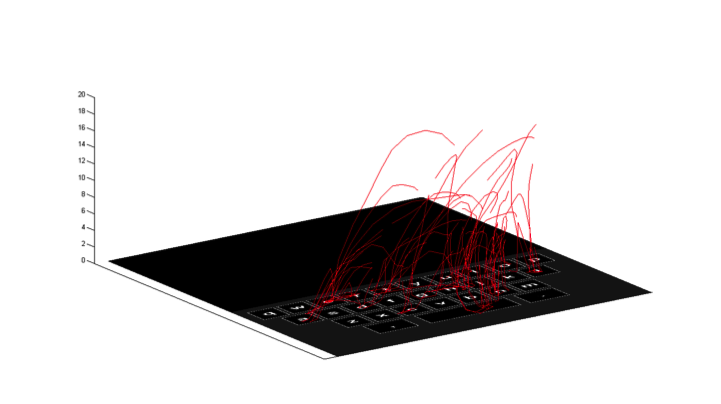}}
\caption{Position of finger typing on mobile phone.}
\label{fig:KeyBoardTraj}
\end{figure}

The ability to track finger position and pose around the screen can allow us to learn more about the range of behaviour in current interactions. This can be applied retrospectively for laboratory analysis of touch interactions (e.g. detailed analysis of typing behaviour with retrospective analysis of finger movements). For example, Figure~\ref{fig:KeyBoardTraj} shows the trace of a finger movement during typing. As this is built into an actual mobile phone, it is now feasible to perform instrumented tests in the wild, rather than requiring a lab-based motion-capture system.

\subsubsection{Design implications}
How can this new ability be used to create new or better interactive systems? The ability to sense finger pose can lead to more accurate and robust models for everyday tasks such as button pushing, gestures and text entry, and could make a difference when the device is subject to disturbances. The improved capacitive sensing approach can also enable novel interaction techniques which depend on richer sensing of finger pose above the device.

Accuracy can be improved by performing inference on the data from multiple frames over time. However, a general caveat for designing interaction around this system is that in typical everyday use, there are many disturbing factors such as device hold, other fingers, different sizes of finger, etc. Furthermore, this paper develops solutions for single digit interaction. While the generalisation to multi-touch is possible, this will require significantly more training data, and is a subject for future exploration. Similarly, the next stage of testing would include acquisition of disturbances such as variability induced by hands gripping the device. 

%A customised model which took into account the covariance between the position and pose and sensors in a causal fashion would be the way to go forward, if the single finger approach is seen as a critical interaction experience.

\newpage
\bibliographystyle{ACM-Reference-Format}
\bibliography{floatbib,touch,Mendeley}


\begin{thebibliography}{41}

%%% the Web address.

\ifx \showCODEN    \undefined \def \showCODEN     #1{\unskip}     \fi
\ifx \showDOI      \undefined \def \showDOI       #1{#1}\fi
\ifx \showISBNx    \undefined \def \showISBNx     #1{\unskip}     \fi
\ifx \showISBNxiii \undefined \def \showISBNxiii  #1{\unskip}     \fi
\ifx \showISSN     \undefined \def \showISSN      #1{\unskip}     \fi
\ifx \showLCCN     \undefined \def \showLCCN      #1{\unskip}     \fi
\ifx \shownote     \undefined \def \shownote      #1{#1}          \fi
\ifx \showarticletitle \undefined \def \showarticletitle #1{#1}   \fi
\ifx \showURL      \undefined \def \showURL       {\relax}        \fi
% invisible to TeX
\providecommand\bibfield[2]{#2}
\providecommand\bibinfo[2]{#2}
\providecommand\natexlab[1]{#1}
\providecommand\showeprint[2][]{arXiv:#2}

\bibitem[\protect\citeauthoryear{Abadi, Agarwal, Barham, Brevdo, Chen, Citro,
  Corrado, Davis, Dean, Devin, Ghemawat, Goodfellow, Harp, Irving, Isard, Jia,
  Jozefowicz, Kaiser, Kudlur, Levenberg, Man\'{e}, Monga, Moore, Murray, Olah,
  Schuster, Shlens, Steiner, Sutskever, Talwar, Tucker, Vanhoucke, Vasudevan,
  Vi\'{e}gas, Vinyals, Warden, Wattenberg, Wicke, Yu, and Zheng}{Abadi
  et~al\mbox{.}}{2015}]%
        {tensorflow2015-whitepaper}
\bibfield{author}{\bibinfo{person}{Mart\'{\i}n Abadi}, \bibinfo{person}{Ashish
  Agarwal}, \bibinfo{person}{Paul Barham}, \bibinfo{person}{Eugene Brevdo},
  \bibinfo{person}{Zhifeng Chen}, \bibinfo{person}{Craig Citro},
  \bibinfo{person}{Greg~S. Corrado}, \bibinfo{person}{Andy Davis},
  \bibinfo{person}{Jeffrey Dean}, \bibinfo{person}{Matthieu Devin},
  \bibinfo{person}{Sanjay Ghemawat}, \bibinfo{person}{Ian Goodfellow},
  \bibinfo{person}{Andrew Harp}, \bibinfo{person}{Geoffrey Irving},
  \bibinfo{person}{Michael Isard}, \bibinfo{person}{Yangqing Jia},
  \bibinfo{person}{Rafal Jozefowicz}, \bibinfo{person}{Lukasz Kaiser},
  \bibinfo{person}{Manjunath Kudlur}, \bibinfo{person}{Josh Levenberg},
  \bibinfo{person}{Dandelion Man\'{e}}, \bibinfo{person}{Rajat Monga},
  \bibinfo{person}{Sherry Moore}, \bibinfo{person}{Derek Murray},
  \bibinfo{person}{Chris Olah}, \bibinfo{person}{Mike Schuster},
  \bibinfo{person}{Jonathon Shlens}, \bibinfo{person}{Benoit Steiner},
  \bibinfo{person}{Ilya Sutskever}, \bibinfo{person}{Kunal Talwar},
  \bibinfo{person}{Paul Tucker}, \bibinfo{person}{Vincent Vanhoucke},
  \bibinfo{person}{Vijay Vasudevan}, \bibinfo{person}{Fernanda Vi\'{e}gas},
  \bibinfo{person}{Oriol Vinyals}, \bibinfo{person}{Pete Warden},
  \bibinfo{person}{Martin Wattenberg}, \bibinfo{person}{Martin Wicke},
  \bibinfo{person}{Yuan Yu}, {and} \bibinfo{person}{Xiaoqiang Zheng}.}
  \bibinfo{year}{2015}\natexlab{}.
\newblock \bibinfo{title}{{TensorFlow}: Large-Scale Machine Learning on
  Heterogeneous Systems}.
\newblock
\newblock
\urldef\tempurl%
\url{https://www.tensorflow.org/}
\showURL{%
\tempurl}
\newblock
\shownote{Software available from tensorflow.org.}


\bibitem[\protect\citeauthoryear{Bachynskyi}{Bachynskyi}{2016}]%
        {Bac16}
\bibfield{author}{\bibinfo{person}{Myroslav Bachynskyi}.}
  \bibinfo{year}{2016}\natexlab{}.
\newblock \emph{\bibinfo{title}{Biomechanical models for human-computer
  interaction}}.
\newblock \bibinfo{thesistype}{Ph.D. Dissertation}.
  \bibinfo{school}{Universit{\"a}t des Saarlandes Saarbr{\"u}cken}.
\newblock


\bibitem[\protect\citeauthoryear{Bachynskyi, Palmas, Oulasvirta, Steimle, and
  Weinkauf}{Bachynskyi et~al\mbox{.}}{2015}]%
        {BacPalOul15}
\bibfield{author}{\bibinfo{person}{Myroslav Bachynskyi},
  \bibinfo{person}{Gregorio Palmas}, \bibinfo{person}{Antti Oulasvirta},
  \bibinfo{person}{J{\"u}rgen Steimle}, {and} \bibinfo{person}{Tino Weinkauf}.}
  \bibinfo{year}{2015}\natexlab{}.
\newblock \showarticletitle{Performance and Ergonomics of Touch Surfaces: A
  Comparative Study using Biomechanical Simulation}. In
  \bibinfo{booktitle}{\emph{Proceedings of the 33rd Annual ACM Conference on
  Human Factors in Computing Systems}}. ACM, \bibinfo{pages}{1817--1826}.
\newblock


\bibitem[\protect\citeauthoryear{Black and Jepson}{Black and Jepson}{1998}]%
        {BlaJep98}
\bibfield{author}{\bibinfo{person}{M.\/~J.\/ Black} {and}
  \bibinfo{person}{A.\/~D.\/ Jepson}.} \bibinfo{year}{1998}\natexlab{}.
\newblock \showarticletitle{{A probabilistic framework for matching temporal
  trajectories: {\{}Condensation{\}}-based recognition of gestures and
  expressions}}. In \bibinfo{booktitle}{\emph{European Conf.{\textbackslash}/
  on Computer Vision, ECCV-98}} \emph{(\bibinfo{series}{LNCS-Series})},
  \bibfield{editor}{\bibinfo{person}{H.\/ Burkhardt} {and}
  \bibinfo{person}{B.\/ Neumann}} (Eds.), Vol.~\bibinfo{volume}{1406}.
  \bibinfo{publisher}{Springer-Verlag}, \bibinfo{address}{Freiburg, Germany},
  \bibinfo{pages}{909--924}.
\newblock


\bibitem[\protect\citeauthoryear{Blake and Isard}{Blake and Isard}{1997}]%
        {BlaIsa97}
\bibfield{author}{\bibinfo{person}{Andrew Blake} {and} \bibinfo{person}{Michael
  Isard}.} \bibinfo{year}{1997}\natexlab{}.
\newblock \showarticletitle{The condensation algorithm-conditional density
  propagation and applications to visual tracking}. In
  \bibinfo{booktitle}{\emph{Advances in Neural Information Processing
  Systems}}. \bibinfo{pages}{361--367}.
\newblock


\bibitem[\protect\citeauthoryear{Conti, Gosling, Oakley, and {O'Hagan}}{Conti
  et~al\mbox{.}}{2009}]%
        {ConGosOak09}
\bibfield{author}{\bibinfo{person}{Stefano Conti}, \bibinfo{person}{John~Paul
  Gosling}, \bibinfo{person}{Jeremy~E. Oakley}, {and} \bibinfo{person}{Anthony
  {O'Hagan}}.} \bibinfo{year}{2009}\natexlab{}.
\newblock \showarticletitle{Gaussian process emulation of dynamic computer
  codes}.
\newblock \bibinfo{journal}{\emph{Biometrika}} (\bibinfo{year}{2009}),
  \bibinfo{pages}{asp028}.
\newblock


\bibitem[\protect\citeauthoryear{Delp, Anderson, Arnold, Loan, Habib, John,
  Guendelman, and Thelen}{Delp et~al\mbox{.}}{2007}]%
        {DelAndArn07}
\bibfield{author}{\bibinfo{person}{Scott~L Delp}, \bibinfo{person}{Frank~C
  Anderson}, \bibinfo{person}{Allison~S Arnold}, \bibinfo{person}{Peter Loan},
  \bibinfo{person}{Ayman Habib}, \bibinfo{person}{Chand~T John},
  \bibinfo{person}{Eran Guendelman}, {and} \bibinfo{person}{Darryl~G Thelen}.}
  \bibinfo{year}{2007}\natexlab{}.
\newblock \showarticletitle{OpenSim: open-source software to create and analyze
  dynamic simulations of movement}.
\newblock \bibinfo{journal}{\emph{IEEE transactions on biomedical engineering}}
  \bibinfo{volume}{54}, \bibinfo{number}{11} (\bibinfo{year}{2007}),
  \bibinfo{pages}{1940--1950}.
\newblock


\bibitem[\protect\citeauthoryear{Geuzaine and Remacle}{Geuzaine and
  Remacle}{2009}]%
        {GeuRem09}
\bibfield{author}{\bibinfo{person}{C. Geuzaine} {and} \bibinfo{person}{J.-F.
  Remacle}.} \bibinfo{year}{2009}\natexlab{}.
\newblock \showarticletitle{{Gmsh}: a three-dimensional finite element mesh
  generator with built-in pre- and post-processing facilities}.
\newblock \bibinfo{journal}{\emph{Internat. J. Numer. Methods Engrg.}}
  \bibinfo{volume}{9}, \bibinfo{number}{11} (\bibinfo{year}{2009}),
  \bibinfo{pages}{1309--1331}.
\newblock


\bibitem[\protect\citeauthoryear{Goodfellow, Bengio, and Courville}{Goodfellow
  et~al\mbox{.}}{2016}]%
        {GooBenCou16}
\bibfield{author}{\bibinfo{person}{Ian Goodfellow}, \bibinfo{person}{Yoshua
  Bengio}, {and} \bibinfo{person}{Aaron Courville}.}
  \bibinfo{year}{2016}\natexlab{}.
\newblock \bibinfo{booktitle}{\emph{Deep learning}}.
\newblock \bibinfo{publisher}{MIT Press}.
\newblock


\bibitem[\protect\citeauthoryear{Grosse-Puppendahl, Braun, Kamieth, and
  Kuijper}{Grosse-Puppendahl et~al\mbox{.}}{2013}]%
        {GroBraKam13}
\bibfield{author}{\bibinfo{person}{Tobias Grosse-Puppendahl},
  \bibinfo{person}{Andreas Braun}, \bibinfo{person}{Felix Kamieth}, {and}
  \bibinfo{person}{Arjan Kuijper}.} \bibinfo{year}{2013}\natexlab{}.
\newblock \showarticletitle{Swiss-cheese Extended: An Object Recognition Method
  for Ubiquitous Interfaces Based on Capacitive Proximity Sensing}. In
  \bibinfo{booktitle}{\emph{Proc. of the SIGCHI Conference on Human Factors in
  Computing Systems}} \emph{(\bibinfo{series}{CHI '13})}.
  \bibinfo{publisher}{ACM}, \bibinfo{pages}{1401--1410}.
\newblock
\urldef\tempurl%
\url{http://doi.acm.org/10.1145/2470654.2466186}
\showURL{%
\tempurl}


\bibitem[\protect\citeauthoryear{Grosse-Puppendahl, Holz, Cohn, Wimmer,
  Bechtold, Hodges, Reynolds, and Smith}{Grosse-Puppendahl
  et~al\mbox{.}}{2017}]%
        {Grosse-puppendahl2017}
\bibfield{author}{\bibinfo{person}{Tobias Grosse-Puppendahl},
  \bibinfo{person}{Christian Holz}, \bibinfo{person}{Gabe Cohn},
  \bibinfo{person}{Raphael Wimmer}, \bibinfo{person}{Oskar Bechtold},
  \bibinfo{person}{Steve Hodges}, \bibinfo{person}{Matthew~S. Reynolds}, {and}
  \bibinfo{person}{Joshua~R. Smith}.} \bibinfo{year}{2017}\natexlab{}.
\newblock \showarticletitle{{Finding Common Ground : A Survey of Capacitive
  Sensing in Human-Computer Interaction}}. In \bibinfo{booktitle}{\emph{ACM SIG
  CHI}}.
\newblock
\showISBNx{9781450346559}
\urldef\tempurl%
\url{https://doi.org/10.1145/3025453.3025808}
\showDOI{\tempurl}


\bibitem[\protect\citeauthoryear{Hester, Brown, Husband, Iliescu, Pruett,
  Summers, and Coleman}{Hester et~al\mbox{.}}{2011}]%
        {HesBroHus11}
\bibfield{author}{\bibinfo{person}{Robert Hester}, \bibinfo{person}{Alison
  Brown}, \bibinfo{person}{Leland Husband}, \bibinfo{person}{Radu Iliescu},
  \bibinfo{person}{William~Andrew Pruett}, \bibinfo{person}{Richard~L.
  Summers}, {and} \bibinfo{person}{Thomas Coleman}.}
  \bibinfo{year}{2011}\natexlab{}.
\newblock \showarticletitle{{HumMod:} a modeling environment for the simulation
  of integrative human physiology}.
\newblock \bibinfo{journal}{\emph{Frontiers in physiology}}
  \bibinfo{volume}{2} (\bibinfo{year}{2011}), \bibinfo{pages}{12}.
\newblock


\bibitem[\protect\citeauthoryear{Hinckley, Heo, Pahud, Holz, Benko, Sellen,
  Banks, Hara, Smyth, and Buxton}{Hinckley et~al\mbox{.}}{2016}]%
        {Hinckley2016}
\bibfield{author}{\bibinfo{person}{Ken Hinckley}, \bibinfo{person}{Seongkook
  Heo}, \bibinfo{person}{Michel Pahud}, \bibinfo{person}{Christian Holz},
  \bibinfo{person}{Hrvoje Benko}, \bibinfo{person}{Abigail Sellen},
  \bibinfo{person}{Richard Banks}, \bibinfo{person}{Kenton~O Hara},
  \bibinfo{person}{Gavin Smyth}, {and} \bibinfo{person}{Bill Buxton}.}
  \bibinfo{year}{2016}\natexlab{}.
\newblock \showarticletitle{{Pre-Touch Sensing for Mobile Interaction}}. In
  \bibinfo{booktitle}{\emph{Proceedings of the 2016 CHI Conference on Human
  Factors in Computing Systems}}. \bibinfo{pages}{2869--2881}.
\newblock
\showISBNx{9781450333627}
\urldef\tempurl%
\url{https://doi.org/10.1145/2858036.2858095}
\showDOI{\tempurl}


\bibitem[\protect\citeauthoryear{Kangasr\"{a}\"{a}si\"{o}, Athukorala, Howes,
  Corander, Kaski, and Oulasvirta}{Kangasr\"{a}\"{a}si\"{o}
  et~al\mbox{.}}{2017}]%
        {KanAthHow18}
\bibfield{author}{\bibinfo{person}{Antti Kangasr\"{a}\"{a}si\"{o}},
  \bibinfo{person}{Kumaripaba Athukorala}, \bibinfo{person}{Andrew Howes},
  \bibinfo{person}{Jukka Corander}, \bibinfo{person}{Samuel Kaski}, {and}
  \bibinfo{person}{Antti Oulasvirta}.} \bibinfo{year}{2017}\natexlab{}.
\newblock \showarticletitle{Inferring Cognitive Models from Data Using
  Approximate Bayesian Computation}. In \bibinfo{booktitle}{\emph{Proceedings
  of the 2017 CHI Conference on Human Factors in Computing Systems}}
  \emph{(\bibinfo{series}{CHI '17})}. \bibinfo{publisher}{ACM},
  \bibinfo{address}{New York, NY, USA}, \bibinfo{pages}{1295--1306}.
\newblock
\showISBNx{978-1-4503-4655-9}


\bibitem[\protect\citeauthoryear{Kennedy and {O'Hagan}}{Kennedy and
  {O'Hagan}}{2001}]%
        {KenOHa01}
\bibfield{author}{\bibinfo{person}{Marc~C. Kennedy} {and}
  \bibinfo{person}{Anthony {O'Hagan}}.} \bibinfo{year}{2001}\natexlab{}.
\newblock \showarticletitle{Bayesian calibration of computer models}.
\newblock \bibinfo{journal}{\emph{Journal of the Royal Statistical Society.
  Series B, Statistical Methodology}} (\bibinfo{year}{2001}),
  \bibinfo{pages}{425--464}.
\newblock


\bibitem[\protect\citeauthoryear{Le, Mayer, Bader, and Henze}{Le
  et~al\mbox{.}}{2017}]%
        {LeMayBad17}
\bibfield{author}{\bibinfo{person}{Huy~Viet Le}, \bibinfo{person}{Sven Mayer},
  \bibinfo{person}{Patrick Bader}, {and} \bibinfo{person}{Niels Henze}.}
  \bibinfo{year}{2017}\natexlab{}.
\newblock \showarticletitle{A Smartphone Prototype for Touch Interaction on the
  Whole Device Surface}. In \bibinfo{booktitle}{\emph{Proceedings of the 19th
  International Conference on Human-Computer Interaction with Mobile Devices
  and Services}} \emph{(\bibinfo{series}{MobileHCI '17})}.
  \bibinfo{publisher}{ACM}, \bibinfo{address}{New York, NY, USA}, Article
  \bibinfo{articleno}{100}, \bibinfo{numpages}{8}~pages.
\newblock
\showISBNx{978-1-4503-5075-4}
\urldef\tempurl%
\url{https://doi.org/10.1145/3098279.3122143}
\showDOI{\tempurl}


\bibitem[\protect\citeauthoryear{Le, Mayer, and Henze}{Le
  et~al\mbox{.}}{2018}]%
        {LeMayHen18}
\bibfield{author}{\bibinfo{person}{Huy~Viet Le}, \bibinfo{person}{Sven Mayer},
  {and} \bibinfo{person}{Niels Henze}.} \bibinfo{year}{2018}\natexlab{}.
\newblock \showarticletitle{InfiniTouch: Finger-Aware Interaction on Fully
  Touch Sensitive Smartphones}. In \bibinfo{booktitle}{\emph{Proceedings of the
  31st Annual ACM Symposium on User Interface Software and Technology
  (UIST'18). ACM, New York, NY, USA}}, Vol.~\bibinfo{volume}{13}.
\newblock


\bibitem[\protect\citeauthoryear{Le~Goc, Taylor, Izadi, and Keskin}{Le~Goc
  et~al\mbox{.}}{2014}]%
        {LeGTayIza14}
\bibfield{author}{\bibinfo{person}{Mathieu Le~Goc}, \bibinfo{person}{Stuart
  Taylor}, \bibinfo{person}{Shahram Izadi}, {and} \bibinfo{person}{Cem
  Keskin}.} \bibinfo{year}{2014}\natexlab{}.
\newblock \showarticletitle{A Low-cost Transparent Electric Field Sensor for 3D
  Interaction on Mobile Devices}. In \bibinfo{booktitle}{\emph{Proceedings of
  the SIGCHI Conference on Human Factors in Computing Systems}}
  \emph{(\bibinfo{series}{CHI '14})}. \bibinfo{publisher}{ACM},
  \bibinfo{pages}{3167--3170}.
\newblock


\bibitem[\protect\citeauthoryear{LeCun, Bengio, and Hinton}{LeCun
  et~al\mbox{.}}{2015}]%
        {LeCBenHin15}
\bibfield{author}{\bibinfo{person}{Yann LeCun}, \bibinfo{person}{Yoshua
  Bengio}, {and} \bibinfo{person}{Geoffrey~E. Hinton}.}
  \bibinfo{year}{2015}\natexlab{}.
\newblock \showarticletitle{{Deep learning}}.
\newblock \bibinfo{journal}{\emph{Nature}} \bibinfo{volume}{521},
  \bibinfo{number}{7553} (\bibinfo{year}{2015}), \bibinfo{pages}{436--444}.
\newblock


\bibitem[\protect\citeauthoryear{MacKay}{MacKay}{2003}]%
        {mackay2003information}
\bibfield{author}{\bibinfo{person}{David~J.C. MacKay}.}
  \bibinfo{year}{2003}\natexlab{}.
\newblock \bibinfo{booktitle}{\emph{Information theory, inference and learning
  algorithms}}.
\newblock \bibinfo{publisher}{Cambridge University Press}.
\newblock


\bibitem[\protect\citeauthoryear{Mansouri and Reinbolt}{Mansouri and
  Reinbolt}{2012}]%
        {ManRei12}
\bibfield{author}{\bibinfo{person}{Misagh Mansouri} {and}
  \bibinfo{person}{Jeffrey~A Reinbolt}.} \bibinfo{year}{2012}\natexlab{}.
\newblock \showarticletitle{A platform for dynamic simulation and control of
  movement based on OpenSim and MATLAB}.
\newblock \bibinfo{journal}{\emph{Journal of biomechanics}}
  \bibinfo{volume}{45}, \bibinfo{number}{8} (\bibinfo{year}{2012}),
  \bibinfo{pages}{1517--1521}.
\newblock


\bibitem[\protect\citeauthoryear{Mayer, Le, and Henze}{Mayer
  et~al\mbox{.}}{2017}]%
        {MayLeHen17}
\bibfield{author}{\bibinfo{person}{Sven Mayer}, \bibinfo{person}{Huy~Viet Le},
  {and} \bibinfo{person}{Niels Henze}.} \bibinfo{year}{2017}\natexlab{}.
\newblock \showarticletitle{Estimating the Finger Orientation on Capacitive
  Touchscreens Using Convolutional Neural Networks}. In
  \bibinfo{booktitle}{\emph{Proceedings of the 2017 ACM International
  Conference on Interactive Surfaces and Spaces}} \emph{(\bibinfo{series}{ISS
  '17})}. \bibinfo{publisher}{ACM}, \bibinfo{address}{New York, NY, USA},
  \bibinfo{pages}{220--229}.
\newblock
\showISBNx{978-1-4503-4691-7}
\urldef\tempurl%
\url{https://doi.org/10.1145/3132272.3134130}
\showDOI{\tempurl}


\bibitem[\protect\citeauthoryear{Mayer, Le, and Henze}{Mayer
  et~al\mbox{.}}{2018}]%
        {MayLeHen18}
\bibfield{author}{\bibinfo{person}{Sven Mayer}, \bibinfo{person}{Huy~Viet Le},
  {and} \bibinfo{person}{Niels Henze}.} \bibinfo{year}{2018}\natexlab{}.
\newblock \showarticletitle{Designing Finger Orientation Input for Mobile
  Touchscreens}. In \bibinfo{booktitle}{\emph{MobileHCI'18}}.
\newblock


\bibitem[\protect\citeauthoryear{Murray-Smith}{Murray-Smith}{2017}]%
        {Mur17}
\bibfield{author}{\bibinfo{person}{R. Murray-Smith}.}
  \bibinfo{year}{2017}\natexlab{}.
\newblock \showarticletitle{Stratified, computational interaction via machine
  learning}. In \bibinfo{booktitle}{\emph{The 18th Yale Workshop on Adaptive
  and Learning Systems}}, \bibfield{editor}{\bibinfo{person}{K.~Narendra}}
  (Ed.).
\newblock


\bibitem[\protect\citeauthoryear{Nguyen, Clune, Bengio, Dosovitskiy, and
  Yosinski}{Nguyen et~al\mbox{.}}{2017}]%
        {PLUG}
\bibfield{author}{\bibinfo{person}{Anh Nguyen}, \bibinfo{person}{Jeff Clune},
  \bibinfo{person}{Yoshua Bengio}, \bibinfo{person}{Alexey Dosovitskiy}, {and}
  \bibinfo{person}{Jason Yosinski}.} \bibinfo{year}{2017}\natexlab{}.
\newblock \showarticletitle{Plug \& Play Generative Networks: Conditional
  Iterative Generation of Images in Latent Space.}. In
  \bibinfo{booktitle}{\emph{CVPR}}, Vol.~\bibinfo{volume}{2}.
  \bibinfo{pages}{7}.
\newblock


\bibitem[\protect\citeauthoryear{O'Hagan}{O'Hagan}{2004}]%
        {Oha04}
\bibfield{author}{\bibinfo{person}{Tony O'Hagan}.}
  \bibinfo{year}{2004}\natexlab{}.
\newblock \showarticletitle{Dicing with the unknown}.
\newblock \bibinfo{journal}{\emph{Significance}} \bibinfo{volume}{1},
  \bibinfo{number}{3} (\bibinfo{year}{2004}), \bibinfo{pages}{132--133}.
\newblock


\bibitem[\protect\citeauthoryear{Oulasvirta, Kim, and Lee}{Oulasvirta
  et~al\mbox{.}}{2018}]%
        {Oul18a}
\bibfield{author}{\bibinfo{person}{Antti Oulasvirta}, \bibinfo{person}{Sunjun
  Kim}, {and} \bibinfo{person}{Byungjoo Lee}.} \bibinfo{year}{2018}\natexlab{}.
\newblock \showarticletitle{Neuromechanics of a Button Press}. In
  \bibinfo{booktitle}{\emph{Proceedings of the 2018 CHI Conference on Human
  Factors in Computing Systems}}. ACM, \bibinfo{pages}{508}.
\newblock


\bibitem[\protect\citeauthoryear{Peherstorfer, Willcox, and
  Gunzburger}{Peherstorfer et~al\mbox{.}}{2018}]%
        {MF}
\bibfield{author}{\bibinfo{person}{Benjamin Peherstorfer},
  \bibinfo{person}{Karen Willcox}, {and} \bibinfo{person}{Max Gunzburger}.}
  \bibinfo{year}{2018}\natexlab{}.
\newblock \showarticletitle{Survey of multifidelity methods in uncertainty
  propagation, inference, and optimization}.
\newblock \bibinfo{journal}{\emph{SIAM Rev.}} \bibinfo{volume}{60},
  \bibinfo{number}{3} (\bibinfo{year}{2018}), \bibinfo{pages}{550--591}.
\newblock


\bibitem[\protect\citeauthoryear{Rogers, Williamson, Stewart, and
  Murray-Smith}{Rogers et~al\mbox{.}}{2010}]%
        {Rogers2010}
\bibfield{author}{\bibinfo{person}{Simon Rogers}, \bibinfo{person}{John
  Williamson}, \bibinfo{person}{Craig Stewart}, {and} \bibinfo{person}{Roderick
  Murray-Smith}.} \bibinfo{year}{2010}\natexlab{}.
\newblock \showarticletitle{FingerCloud: Uncertainty and Autonomy Handover in
  Capacitive Sensing}. In \bibinfo{booktitle}{\emph{Proceedings of the SIGCHI
  Conference on Human Factors in Computing Systems}}
  \emph{(\bibinfo{series}{CHI '10})}. \bibinfo{publisher}{ACM},
  \bibinfo{address}{New York, NY, USA}, \bibinfo{pages}{577--580}.
\newblock
\showISBNx{978-1-60558-929-9}
\urldef\tempurl%
\url{https://doi.org/10.1145/1753326.1753412}
\showDOI{\tempurl}


\bibitem[\protect\citeauthoryear{Rogers, Williamson, Stewart, and
  Murray-Smith}{Rogers et~al\mbox{.}}{2011a}]%
        {RogWilSte11}
\bibfield{author}{\bibinfo{person}{Simon Rogers}, \bibinfo{person}{John
  Williamson}, \bibinfo{person}{Craig Stewart}, {and} \bibinfo{person}{Roderick
  Murray-Smith}.} \bibinfo{year}{2011}\natexlab{a}.
\newblock \showarticletitle{AnglePose: robust, precise capacitive touch
  tracking via 3D orientation estimation}. In
  \bibinfo{booktitle}{\emph{Proceedings of the SIGCHI Conference on Human
  Factors in Computing Systems}}. ACM, \bibinfo{pages}{2575--2584}.
\newblock


\bibitem[\protect\citeauthoryear{Rogers, Williamson, Stewart, and
  Murray-Smith}{Rogers et~al\mbox{.}}{2011b}]%
        {Rogers2011a}
\bibfield{author}{\bibinfo{person}{Simon Rogers}, \bibinfo{person}{John
  Williamson}, \bibinfo{person}{Craig Stewart}, {and} \bibinfo{person}{Roderick
  Murray-Smith}.} \bibinfo{year}{2011}\natexlab{b}.
\newblock \showarticletitle{AnglePose: Robust, Precise Capacitive Touch
  Tracking via 3D Orientation Estimation}. In
  \bibinfo{booktitle}{\emph{Proceedings of the SIGCHI Conference on Human
  Factors in Computing Systems}} \emph{(\bibinfo{series}{CHI '11})}.
  \bibinfo{publisher}{ACM}, \bibinfo{address}{New York, NY, USA},
  \bibinfo{pages}{2575--2584}.
\newblock
\showISBNx{978-1-4503-0228-9}
\urldef\tempurl%
\url{https://doi.org/10.1145/1978942.1979318}
\showDOI{\tempurl}


\bibitem[\protect\citeauthoryear{Sacks, Welch, Mitchell, and Wynn}{Sacks
  et~al\mbox{.}}{1989}]%
        {SacWelTob89}
\bibfield{author}{\bibinfo{person}{Jerome Sacks}, \bibinfo{person}{William~J
  Welch}, \bibinfo{person}{Toby~J Mitchell}, {and} \bibinfo{person}{Henry~P
  Wynn}.} \bibinfo{year}{1989}\natexlab{}.
\newblock \showarticletitle{[Design and Analysis of Computer Experiments]:
  Rejoinder}.
\newblock \bibinfo{journal}{\emph{Statist. Sci.}} (\bibinfo{year}{1989}),
  \bibinfo{pages}{433--435}.
\newblock


\bibitem[\protect\citeauthoryear{S{\"a}rkk{\"a}}{S{\"a}rkk{\"a}}{2013}]%
        {sarkka2013bayesian}
\bibfield{author}{\bibinfo{person}{Simo S{\"a}rkk{\"a}}.}
  \bibinfo{year}{2013}\natexlab{}.
\newblock \bibinfo{booktitle}{\emph{Bayesian filtering and smoothing}}.
  Vol.~\bibinfo{volume}{3}.
\newblock \bibinfo{publisher}{Cambridge University Press}.
\newblock


\bibitem[\protect\citeauthoryear{Schmidhuber}{Schmidhuber}{2015}]%
        {Sch15}
\bibfield{author}{\bibinfo{person}{J{\"u}rgen Schmidhuber}.}
  \bibinfo{year}{2015}\natexlab{}.
\newblock \showarticletitle{Deep learning in neural networks: An overview}.
\newblock \bibinfo{journal}{\emph{Neural Networks}}  \bibinfo{volume}{61}
  (\bibinfo{year}{2015}), \bibinfo{pages}{85--117}.
\newblock


\bibitem[\protect\citeauthoryear{Smith}{Smith}{1996}]%
        {Smi96}
\bibfield{author}{\bibinfo{person}{J.~R. Smith}.}
  \bibinfo{year}{1996}\natexlab{}.
\newblock \showarticletitle{Field mice: Extracting hand geometry from electric
  field measurements}.
\newblock \bibinfo{journal}{\emph{IBM Systems Journal}} \bibinfo{volume}{35},
  \bibinfo{number}{3.4} (\bibinfo{year}{1996}), \bibinfo{pages}{587--608}.
\newblock


\bibitem[\protect\citeauthoryear{Sohn, Lee, and Yan}{Sohn
  et~al\mbox{.}}{2015}]%
        {CVAE}
\bibfield{author}{\bibinfo{person}{Kihyuk Sohn}, \bibinfo{person}{Honglak Lee},
  {and} \bibinfo{person}{Xinchen Yan}.} \bibinfo{year}{2015}\natexlab{}.
\newblock \showarticletitle{Learning structured output representation using
  deep conditional generative models}. In \bibinfo{booktitle}{\emph{Advances in
  Neural Information Processing Systems}}. \bibinfo{pages}{3483--3491}.
\newblock


\bibitem[\protect\citeauthoryear{Tarantola}{Tarantola}{2005}]%
        {Tar05}
\bibfield{author}{\bibinfo{person}{Albert Tarantola}.}
  \bibinfo{year}{2005}\natexlab{}.
\newblock \bibinfo{booktitle}{\emph{Inverse problem theory and methods for
  model parameter estimation}}. Vol.~\bibinfo{volume}{89}.
\newblock \bibinfo{publisher}{{SIAM}}.
\newblock


\bibitem[\protect\citeauthoryear{Tonolini, Radford, Turpin, Faccio, and
  Murray-Smith}{Tonolini et~al\mbox{.}}{2020}]%
        {TonRadTur20}
\bibfield{author}{\bibinfo{person}{Francesco Tonolini}, \bibinfo{person}{Jack
  Radford}, \bibinfo{person}{Alex Turpin}, \bibinfo{person}{Daniele Faccio},
  {and} \bibinfo{person}{Roderick Murray-Smith}.}
  \bibinfo{year}{2020}\natexlab{}.
\newblock \showarticletitle{Variational Inference for Computational Imaging
  Inverse Problems}.
\newblock \bibinfo{journal}{\emph{Journal of Machine Learning Research}}
  \bibinfo{volume}{21}, \bibinfo{number}{179} (\bibinfo{year}{2020}),
  \bibinfo{pages}{1--46}.
\newblock
\urldef\tempurl%
\url{http://jmlr.org/papers/v21/20-151.html}
\showURL{%
\tempurl}


\bibitem[\protect\citeauthoryear{Turkman, Paulino, and M{\"u}ller}{Turkman
  et~al\mbox{.}}{2019}]%
        {TurPauMul19}
\bibfield{author}{\bibinfo{person}{M.~Ant{\'o}nia~Amaral Turkman},
  \bibinfo{person}{Carlos~Daniel Paulino}, {and} \bibinfo{person}{Peter
  M{\"u}ller}.} \bibinfo{year}{2019}\natexlab{}.
\newblock \bibinfo{booktitle}{\emph{Computational Bayesian Statistics: An
  Introduction}}. Vol.~\bibinfo{volume}{11}.
\newblock \bibinfo{publisher}{Cambridge University Press}.
\newblock


\bibitem[\protect\citeauthoryear{Williamson}{Williamson}{2016}]%
        {Wil16}
\bibfield{author}{\bibinfo{person}{John Williamson}.}
  \bibinfo{year}{2016}\natexlab{}.
\newblock \showarticletitle{{Fingers of a hand oscillate together}}. In
  \bibinfo{booktitle}{\emph{ACM SIG CHI, San Jose}}. \bibinfo{publisher}{ACM},
  \bibinfo{pages}{3433--3437}.
\newblock
\urldef\tempurl%
\url{https://doi.org/10.1145/2858036.2858235}
\showDOI{\tempurl}


\bibitem[\protect\citeauthoryear{Xiao, Schwarz, and Harrison}{Xiao
  et~al\mbox{.}}{2015}]%
        {XiaSchHar15}
\bibfield{author}{\bibinfo{person}{Robert Xiao}, \bibinfo{person}{Julia
  Schwarz}, {and} \bibinfo{person}{Chris Harrison}.}
  \bibinfo{year}{2015}\natexlab{}.
\newblock \showarticletitle{Estimating 3D Finger Angle on Commodity
  Touchscreens}. In \bibinfo{booktitle}{\emph{Proceedings of the 2015
  International Conference on Interactive Tabletops \& Surfaces}}
  \emph{(\bibinfo{series}{ITS '15})}. \bibinfo{publisher}{ACM},
  \bibinfo{address}{New York, NY, USA}, \bibinfo{pages}{47--50}.
\newblock
\showISBNx{978-1-4503-3899-8}
\urldef\tempurl%
\url{https://doi.org/10.1145/2817721.2817737}
\showDOI{\tempurl}


\end{thebibliography}

\end{document}